\documentclass[12pt]{article}

\usepackage{scicite}

\usepackage{graphicx}
\usepackage{color}
\usepackage{epsf}
\usepackage{times}

\newenvironment{sciabstract}{%
\begin{quote} \bf}
{\end{quote}}

\newcounter{lastnote}

\title{Observation of small Fermi pockets protected by clean CuO$_2$ sheets of a high-$T_c$ superconductor}

\author
{So~Kunisada,$^{{1}}$ 
Shunsuke~Isono,$^{2}$
Yoshimitsu~Kohama,$^{1}$ \\
Shiro~Sakai,$^{3}$ 
C\'edric Bareille,$^{1}$ 
Shunsuke~Sakuragi,$^{1}$ 
Ryo~Noguchi,$^{1}$ \\
Kifu~Kurokawa,$^{1}$ 
Kenta~Kuroda,$^{1}$ 
Yukiaki~Ishida,$^{1}$ 
Shintaro~Adachi,$^{4}$\\
Ryotaro~Sekine,$^{2}$
Timur~K.~Kim,$^{5}$
Cephise~Cacho,$^{5}$
Shik~Shin,$^{1,7}$\\
Takami~Tohyama,$^{6}$
Kazuyasu~Tokiwa,$^{2\ast}$
Takeshi~Kondo $^{1,7\ast}$\\
 \\
\normalsize{$^{1}$ISSP, University of Tokyo, Kashiwa, Chiba 277-8581, Japan.}\\
\normalsize{$^{2}$Department of Applied Electronics, Tokyo University of Science, Tokyo 125-8585, Japan.}\\
\normalsize{$^{3}$RIKEN Center for Emergent Matter Science (CEMS), Wako, Saitama 351-0198, Japan.}\\
\normalsize{$^{4}$MANA, National Institute for Materials Science, Tsukuba, Ibaraki 305-0047, Japan.}\\
\normalsize{$^{5}$Diamond Light Source, Harwell Campus, Didcot, OX11 0DE, United Kingdom.}\\
\normalsize{$^{6}$Department of Applied Physics, Tokyo University of Science, Tokyo 125-8585, Japan.}\\
\normalsize{$^{7}$AIST-UTokyo Advanced Operando-Measurement Technology Open Innovation}\\
\normalsize{Laboratory (OPERANDO-OIL), Kashiwa, Chiba 277-8581, Japan}\\
\normalsize{$^\ast$To whom correspondence should be addressed; E-mail: }\\
\normalsize{tokiwa@rs.tus.ac.jp and kondo1215@issp.u-tokyo.ac.jp}
}

\date{}

\begin{document} 


\baselineskip24pt


\maketitle


\begin{sciabstract}
The superconductivity of high transition temperature ($T_c$) occurs in copper oxides with carrier-doping to an antiferromagnetic (AF) Mott insulator \cite{Dagotto,Imada}. This discovery more than thirty years ago immediately led to a prediction about the formation of a small Fermi pocket. This structure, however, has not yet been detected, while it could be a key element in relating high-$T_c$ superconductivity to Mott physics. To address this long-standing issue, we investigate the electronic structure of a five-layered Ba$_2$Ca$_4$Cu$_5$O$_{10}$(F,O)$_2$ with inner CuO$_2$ planes demonstrated to be cleanest ever in cuprates \cite{Mukuda,Shimizu}. Most surprisingly, we find small Fermi surface (FS) pockets closed around ($\pi$/2,$\pi$/2) consistently by angle-resolved photoemission spectroscopy (ARPES) and quantum oscillation measurements.  The $d$-wave  superconducting gap opens along the pocket, revealing the coexistence between the superconductivity and AF ordering in the same CuO$_2$ sheet. Our data further indicate that the superconductivity can occur without contribution from the states near the antinodal region, which are shared by other competing excitations such as the charge density wave (CDW) and pseudogap states \cite{Kondo,Chang,Damascelli_CDW,Wise_CDW,Hashimoto,Ghiringhelli}. This will have significant implications for understanding the superconductivity and puzzling Fermi arc phenomena in cuprates \cite{Norman}.
\end{sciabstract}




In the AF Mott insulating phase of cuprates \cite{Dagotto,Imada}, the top of the occupied band with a convex upward is located at  ($\pi$/2,$\pi$/2) \cite{Wells_half,Ronning_half}. By slightly hole-doping to it, the Fermi energy is expected to cross the lower Hubbard band (or charge transfer band), turning the state to be metallic \cite{Ronning_doping,Tohyama}. Observation of a Fermi pocket by ARPES has been reported for a single-layered Bi$_2$Sr$_2$CuO$_{6+\delta}$ (Bi2201) \cite{Zhou_Nature}.  This pocket was however detected significantly off  ($\pi$/2,$\pi$/2), and only for a specific region relatively close to the optimal doping. This structure was later claimed to be an artifact coming from the distorted crystal structure inherent in this compound \cite{Matsui_modulation}, which generates replicas of the main band in photoemission \cite{King_PRL}. Moreover, the Fermi pocket in Bi2201 has not been detected by the bulk-sensitive quantum oscillation measurements. 

The quantum oscillations capturing small FS pockets in hole-doped cuprates have been observed in three compounds: a double-layered YBa$_2$Cu$_3$O$_{6.5 }$ (Y123) \cite{YBCO_quantum} and YBa$_2$Cu$_4$O$_8$ (Y124) \cite{Y124_quamtum}, and a single-layered HgBa$_2$CuO$_{4+\delta}$ (Hg1201) \cite{Hg_quantum}. However, the corresponding structures have not been detected by ARPES.  The pockets were soon turned out to be negative in charges, and now it has been widely acknowledged that these are not of the genuine ground state, but of reconstructed FSs under an external magnetic field, which eliminates the superconductivity and instead develops the competing charge density wave \cite{YBCO_Negative}. Thus, while long desired since the discovery of cuprates, the experimental observation of the small hole pocket in lightly doped cuprates has been lacking to date.


In the single- and double-layered prototype materials, the superconducting CuO$_2$ sheet is adjacent to dopant layers with random vacancies, which cause spatially inhomogeneous state as revealed by scanning tunneling microscopy \cite{STM}. This undesirable situation is resolved in the triple and more layered systems, where inner CuO$_2$ planes are screened from the disorder of outermost dopant layers \cite{Mukuda,Shimizu}. The NMR data indicate that the protected inner planes are not only structurally flatter but also more homogeneously carrier-doped than the outer planes, and those in five-layered cuprates are much cleaner than the CuO$_2$ planes in Y123 and Hg1201 \cite{Mukuda,Shimizu}.


In this study, we have particularly selected the five-layered Ba$_2$Ca$_4$Cu$_5$O$_{10}$(F,O)$_2$ ($T_c=65$ K) with lightly doped inner planes (Fig. 1C and Supplemental Fig. S1), which is easy to cleave, and hence suitable for ARPES. Significantly, the crystal is free from structural distortion, and therefore from the observation of artificial Fermi pockets. Moreover, the AF-ordered phase extends inside the superconducting region as a common property of the triple and more layered cuprates with interlayer magnetic coupling \cite{Mukuda,Shimizu}, which will bring a firm formation of Fermi pocket. This material, therefore, would be best suited to unveil electronic structure inherent to the lightly doped Mott insulator. On top of it, the protected inner plane of cuprates is an unexplored yet crucial object which should be unveiled, since it is a key platform to achieve the highest $T_c$ among all the existing substances as demonstrated by the trilayer HgBa$_2$Ca$_2$Cu$_3$O$_{8+\delta}$ \cite{Hg1223}. 

Figure 1A plots the ARPES intensities integrated close to the Fermi level ($E_F$) measured at the lowest temperature ($T$ = 5 K) by a laser-ARPES with high energy and momentum resolutions. Surprisingly, we find two hole pockets centered at ($\pi$/2,$\pi$/2), in addition to a Fermi arc typical for underdoped cuprates \cite{Norman}. These are also confirmed in band dispersions (Fig. 1K, 1L, and 1M); the back side of the Fermi pockets becomes clear by changing the color scale of ARPES images (Fig. 1H, 1I, and 1J). The corresponding peaks are also clearly seen in momentum distribution curves (MDCs) at $E_F$ (Fig. 1E, 1F, and 1G), while there is no such signature for the Fermi arc. 

Our high-quality data justify the pockets as intrinsic FSs and rule out other possibilities as follows. The pockets (Fig. 1B) and occupied dispersions (Fig. 1N) both vertically cross the antiferromagnetic zone boundary (AFZB); these structures are exactly what are expected for one band, and not compatible with overlapped two bands (Supplemental Fig. S2A and S2B). Moreover, the band dispersion along the AFZB line exhibits MDCs with very sharp peaks (Fig. 1O and 1P) comparable to those along the nodal cut, which will never be obtained for overlapped bands (Supplemental Fig. S2A). The scenario of surface reconstruction can also be denied from the fact that there is no folding back for the Fermi arc (dashed green arrows in Fig. 1E-1G). While the signature of AF order has been detected from not only the inner planes but also the outer planes by NMR measurements, the former is much stronger, being compatible with our ARPES results showing apparent Fermi pockets only for the inner planes. 

Since the hole carriers are doped by the O substitution for the F sites (Fig. 1C),  outer CuO$_2$ layers close to the BaF dopant-layer are more doped than inner layers  \cite{Iyo,Kotegawa}.  Thus, the smallest pocket, larger pocket, and  Fermi arc should be formed dominantly by the innermost plane (IP$_0$), the second inner planes (IP$_1$), and the outermost planes (OP), respectively (arrows in Fig. 1A). Typically, the spectral peaks get broader with fewer carriers due to a stronger correlation effect as demonstrated for Bi$_2$Sr$_2$CaCu$_2$O$_{8+\delta}$ (Bi2212) (Supplemental Fig. S3); with underdoping, the nodal $k_F$ point gets closer to AFZB, called ``hot-spots", on which the AF fluctuation with a ($\pi$, $\pi$) vector causes significant electron scattering. Interestingly, we find the opposite trend here (Fig. 1D): the peak of nodal spectrum is sharper for the pockets than for the Fermi arc, and it becomes sharpest for the smallest pocket (even sharper than that in optimally doped Bi2212; see Supplemental Fig. S3L).  This indicates that IP$_0$ is indeed very clean and has carriers with long lifetime, while the suppression of AF fluctuations may be another factor that causes less scattering in the inner planes.

It would be crucial to confirm the same Fermi pockets by a bulk sensitive probe. We have hence measured de Haas-van Alphen effect (dHvA) (Fig. 2A and Supplemental Fig. S4B) and Shubnikov-de Haas effect (SdH) (Supplemental Fig. S4A) through torque and contactless resistivity measurements, respectively, and from the both detected quantum oscillations indicating two-dimensional pockets (Supplemental Fig. S5). In the fast Fourier transformation (FFT) spectra for dHvA (Fig. 2B), we find mainly two peaks at F$_0$ = 147 T and F$_1$ =318 T, which correspond to the FS area covering 2.1 $\%$ and 4.5 $\%$ of the Brillouin zone, respectively. Importantly, these values almost perfectly agree to the ARPES results (2.2 $\%$ and 4.7 $\%$).  The charge carriers are found to be always hole-type regardless of temperature with no sign inversion even at high magnetic fields as confirmed by the behavior of Hall resistance against the magnetic field  (see Fig. S6 and related discussion in Supplementary Materials \cite{SM}).Therefore, the Fermi pockets captured by quantum oscillations should be the same as those by ARPES, rather than the reconstructed FSs with electron-type carriers as reported for Y123 and Y124  \cite{YBCO_Negative}.

We have also confirmed a good agreement in the mass of conduction electrons: 0.69 $m_0$ for IP$_0$ and 0.74 $m_0$ for IP$_1$ by dHvA (Supplemental Fig. S4C), and $\sim$0.7 $m_0$ for the two pockets by ARPES. Another agreements with ARPES are seen in the Dingle temperature ($T_D$), proportional to the scattering rate (Supplemental Fig. S4D), and the mean free path ($l$);  the $T_D$ and  $l$ values are estimated to be lower and longer, respectively, for the smaller pocket ($T_D$=6.5 K, $l$=210 $\mathrm{\AA}$) than those for the larger pocket ($T_D$=11.8 K, $l$=160 $\mathrm{\AA}$), which is consistent with ARPES spectra showing sharper peaks in the former (Fig. 1D) and agrees with the argument that IP$_0$ is cleaner than IP$_1$. We also note that these values are comparable to those of Y123 ($T_D$=6.2 K, $l$=200 $\mathrm{\AA}$) \cite{YBCO_quantum,Chan_NC}, and lower and longer, respectively, than those of Hg1201 ($T_D$=18 K, $l$=85 $\mathrm{\AA}$) \cite{Hg_quantum,Chan_NC}, which further verifies that the carriers of protected inner planes have an exceptionally high mobility despite in such a  lightly doped regime. 

The laser-ARPES with low energy photons cannot access the momentum region close to the Brillouin zone boundary. Therefore, we have used synchrotron light to determine the whole band structure. Moreover, the photon energies ($h\nu$) are tunable, thus the band-selective measurement becomes available with the matrix element effect in photoemission \cite{Ideta}. We have fine-tuned the photon energy in reference to previous research for Ca$_{2}$CuO$_2$Cl$_2$ (CCOC), which demonstrated that the spectral intensities for the folded band are sensitive to the used photon energy, and could become even stronger than those of the main band \cite{Ronning_half,CCOC_300K}. We find that $h\nu$ of 70 eV is best suited not only to observe the Fermi pockets selectively but also to enhance the ARPES intensity for the back side of the pockets (Fig. 3A and 3C); our data clearly exhibit a pocket around ($\pi$/2,$\pi$/2), while two pockets are indistinguishable because of limited resolutions in the synchrotron ARPES. 

The energy dispersion along (-$\pi$,$\pi$)-($\pi$,-$\pi$) crossing the pocket (Fig. 3B) contrasts to  a large parabolic dispersion seen at $h\nu=100$ eV (Supplemental Fig. S7B), which selectively observes the Fermi arc band (Supplemental Fig. S7A-S7G). To unveil the whole band shape for the Fermi pockets, we also extracted the ARPES dispersion along AFZB over multiple Brillouin zones (Fig. 3D);  a periodic pattern has been obtained. The band determined by a tight-binding fit to our ARPES data is plotted in Fig. 3E. We find that the saddle point at the zone edge \cite{Damascelli}, a famous feature in cuprates, has totally gone.  Instead, a parabola with convex downward disperses down below -1 eV at ($\pi$,0) (see Fig. 3D). Therefore, the CDW and pseudogap states known to emerge around ($\pi$,0) with the energy scale of $\sim 100$ meV \cite{Kondo,Wise_CDW,Zhou_Bi2201} cannot develop in the inner planes, which lack electrons required to generate these excitations in the band structure; this sharply contrasts to the situation in the outer planes, where the psuedogap opens around ($\pi$,0) as in other underdoped cuprates, and thus the CDW state is likewise expected to occur \cite{Ghiringhelli,Chang}.

The band shape we observed is compatible with that of Mott insulating CCOC  \cite{Ronning_science} and Sr$_{2}$CuO$_2$Cl$_2$ (SCOC)  \cite{Wells_half}, while our sample is metallic, and hence the chemical potential crosses the lower Hubbard band (Fig. 3F). In our sample, the bandwidth ($W$) is about $1.5$ eV and a kink structure is observable around $-0.5$ eV in the band dispersion. Such a waterfall-like dispersion with a large energy width of $1.3\sim2.0$ eV has been commonly observed in many cuprate compounds \cite{Chang_dispersion}, including the insulating CCOC ($W$$\sim1.8$ eV) with the lower Hubbard band fully occupied \cite{Ronning_waterfall}.  The Fermi pockets in the carrier-doped Mott band of our sample seem to persist above $T_N$ ($\sim135$ K), since the temperature dependence of the Hall coefficient (Supplemental Fig. S6C) exhibits no significant variation across $T_N$. This agrees to the ARPES studies of Mott-insulating CCOC ($T_N$=245 K) and SCOC ($T_N$=256 K), where the band folding has been clearly observed at the room temperature \cite{CCOC_300K} or even 100 K higher than $T_N$ \cite{Wells_half}. 

To investigate the superconducting gap \cite{Harter_PRL,Sugimoto_STM,Yoshida_PRL,Matsui_PRL}, we have used a high-resolution laser-ARPES. In Fig. 4A, we plot the energy distribution curves (EDCs) along the Fermi pocket for IP$_1$ (red oval in Fig. 4F) especially for the main side of the Brillouin zone (green circles in Fig. 4G). As expected for a $d$-wave gap structure, the spectral edge reaches $E_F$ with no gap in the diagonal direction ($\phi$=$0^{\circ}$; $\phi$ is defined in Fig. 4G), and it shifts to higher binding energies with a gap opening off the node. This gap behavior is visualized in Fig. 4B with the EDCs symmetrized across $E_F$:  the nodal spectrum with one peak changes to two-peak structure off the node, and the gap increases up to the tip of the ellipsoidal Fermi pocket ($\phi$=$90^{\circ}$).  Significantly, we have obtained almost the same results (Fig. 4C and 4D) for the folded side of the Brillouin zone (purple circles in Fig. 4G), which are summarized in Fig. 4E. 

These results suggest that the superconductivity occurs in the inner CuO$_2$ plane (IP$_1$), not due to the proximity effect from the superconducting outer planes with the Fermi arc. We demonstrate this within a model calculation (Supplemental Fig. S8) by comparing the spectra calculated with and without a superconducting order in IP$_1$. In the latter case, the EDC at the larger Fermi pocket (mainly contributed by IP$_1$) never shows a visible size of superconducting gap unlike the experiment; the gap is reproduced only when IP$_1$ itself is superconducting. In the model, the Fermi pockets are formed by a band-folding due to the AF order in the inner planes, so that the results also demonstrate a microscopic coexistence of the AF order and superconductivity in IP$_1$. We also note that the coexisting of the AF order and superconductivity
in lightly-doped CuO$_2$ plane has been confirmed by numerical calculations 
for the Hubbard model \cite{Coexist_DMFT,Kobayashi_theory,Coexist_Hubbard}.

To validate the relationship between the different layers more directly, we have compared the superconducting gaps on the Fermi pocket and arc measured by ARPES (Supplemental Fig. S9A-S9D), and found that the gap size in the pocket for IP$_1$ is slightly larger than that in the Fermi arc for OP (Supplemental Fig. S9H). The size relationship gives the trend opposite to the expectation from the proximity scenario, which can hence be ruled out. Since the Fermi pocket emerges in the doped Mott band with the AF ordering  (Fig. 3F), the observation of superconducting gap along it (Fig. 4E) directly demonstrates the coexistence of superconductivity and AF order at a CuO$_2$ sheet (Fig. 4H) \cite{Mukuda,Shimizu}; it is striking that such a small amount of carriers ($p \sim 0.04$) form superconducting pairs under the majority background of the AF-ordered state. 

Along the smallest Fermi pocket, the superconducting gap is found to be almost zero within the experimental resolution (Supplemental Fig. S9H). This extreme difference in the superconductivity between the two pockets has two significant implications. First, the two pockets are almost separately contributed by IP$_0$ and IP$_1$, respectively, since otherwise the mixing of layers would produce superconducting gaps of similar magnitudes for both pockets. Secondly, the electronic state of IP$_0$,  less doped than IP$_1$, should be situated outside of the $T_c$ dome in the phase diagram.

The data showing the larger superconducting gap in the Fermi pocket than Fermi arc (Supplemental Fig. S9H) imply that the electron pairing gets more stabilized in the former despite less doped by avoiding the competition with other ordered states, which could develop around the zone edge ($\pi$,0). This may be the main reason why the superconductivity persists down to the close vicinity of half-filled Mott state in the five-layered cuprates,  contrasting to the case of the single layered cuprates with disordered CuO$_2$ layers, which are insulating up to about $10\% $ carrier doping \cite{Zhou_Bi2201}. Nonetheless, further investigations would be necessary to clarify which layer actually trigger the superconductivity in bulk since the relationship between the $T_c$ value and the superconducting gap magnitude in cuprates has been still a debated issue; it is particularly intriguing for the pairing mechanism that larger pairing gaps open in the CuO$_2$ sheet with the AF-fluctuation suppressed.  Our results will be significant not only to solve a long-standing puzzle on the Fermi arc phenomena but also to revisit the Mott physics leading to the electron pairing in cuprates, which has been built up to now based mainly on the research for a single- and double-layered compounds with the inhomogeneous electronic state.

{\bf Acknowledgments:} 
 We thank K. Akiba for supporting the analysis of quantum oscillation,  T. Kiss for useful discussions, 
and T. Yajima for technical assistance in the X-ray measurements, performed using facilities of the Institute for Solid State Physics, the University of Tokyo.  We thank Diamond Light Source for access to beamline I05 under proposal SI20445 and SI20446 that contributed to the results presented here.
This work was supported by the JSPS KAKENHI (Grants No. JP19K03722, JP18H01165, JP16H06345, JP19H01829, and JP19H02683),  by AIST-UTokyo OPERANDO-OIL, and by MEXT Q-LEAP.



\clearpage

\begin{figure} 
\begin{center}
\includegraphics[width=6in]{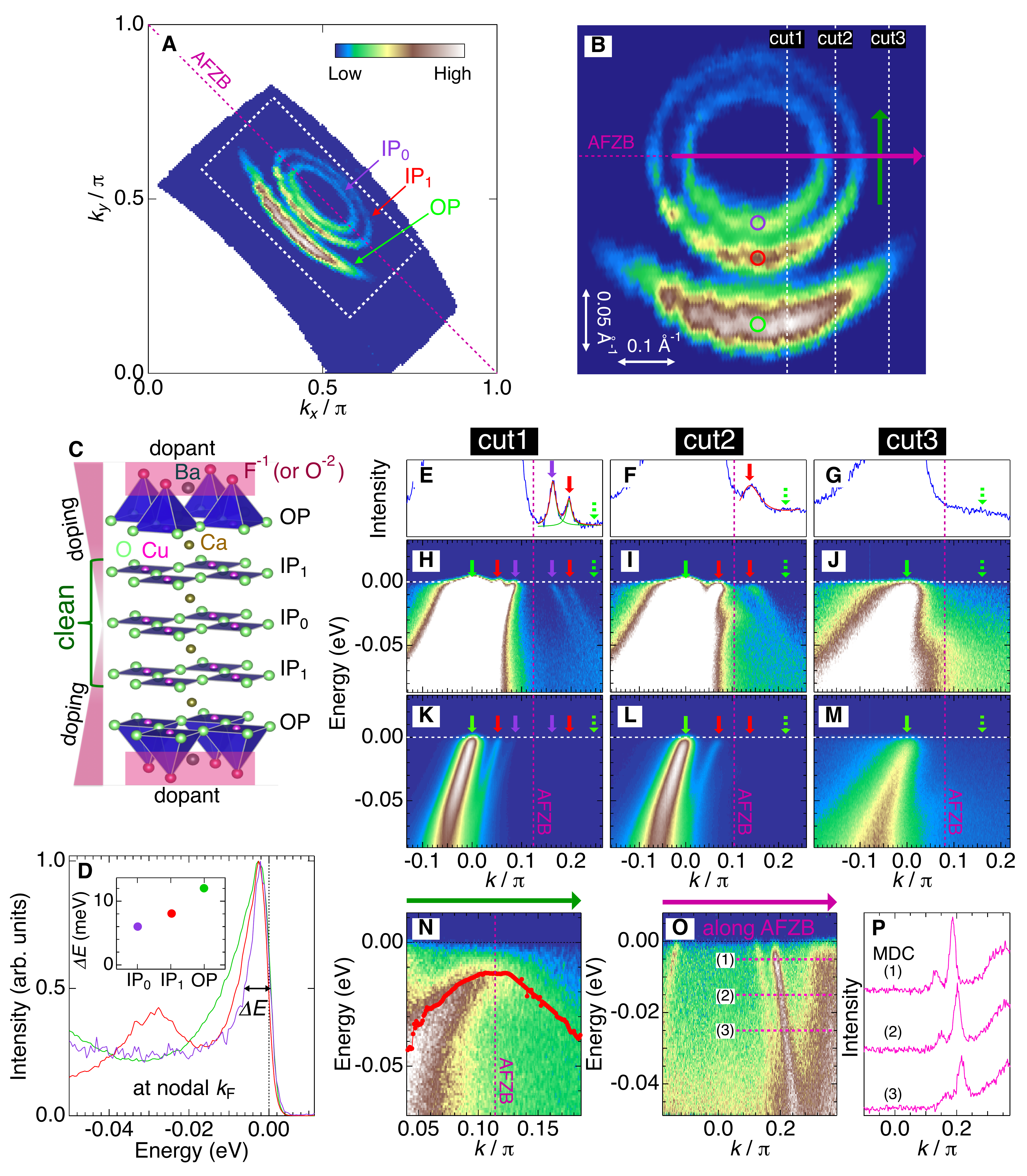}
\end{center}
\label{fig1}
\end{figure}

\clearpage
\newpage

\noindent {\bf Fig. 1.} 
{\bf Small Fermi pockets revealed by laser-ARPES.}
({\bf A}) Fermi surface (FS) mapping obtained by integrating ARPES intensities within an energy window of 10 meV about $E_F$. The arrows point to FSs dominated by the innermost plane (IP$_0$), the second inner planes (IP$_1$), and the outermost planes (OP), which are depicted in (C).
({\bf B}) Magnified image of (A) in the region marked by a white dotted square.  
Dimensions along the horizontal and vertical axes are noted with arrows in the image.
({\bf C}) Five CuO$_2$ sheets in the crystal structure of Ba$_2$Ca$_4$Cu$_5$O$_{10}$(F,O)$_2$. 
({\bf D}) The spectra at the nodes (marked by colored  circles in (B)) for the smaller and larger Fermi pockets (IP$_0$ and IP$_1$, respectively) and Fermi arc (OP). Inset plots the energy width of energy distribution curves ($\Delta E$) for each FS (or each CuO$_2$ sheet). Note that the broad peak around -0.03 eV in the red spectrum comes from the energy state of the band for IP$_0$, which stays energetically below the band of IP$_1$ [for example, see panel (H)]. ({\bf E to G}) Momentum distribution curves (MDCs) at $E_F$ for each panel of (H to J). 
The vertical axis values are decreased to clearly exhibit peaks for the back sides of Fermi pockets; the peaks for the pockets (fitted by Lorentzian curves) are clearly visible, whereas the one corresponding to the Fermi arc is missing (a green dashed arrow). 
({\bf H to J}) The same images as in (K to M), but the color scale is changed to emphasize the folded bands.
({\bf K to M})  ARPES dispersions obtained along the momentum cuts indicated in (B) by white dotted lines (cut1, cut1, and cut3).
({\bf N to O}) ARPES dispersions extracted along the momentum cuts indicated in (B) by green and purple arrows, respectively.  
({\bf P}) MDCs along dashed lines in (O).

\clearpage
\newpage

\begin{figure} 
\begin{center}
\includegraphics[width=5.5in]{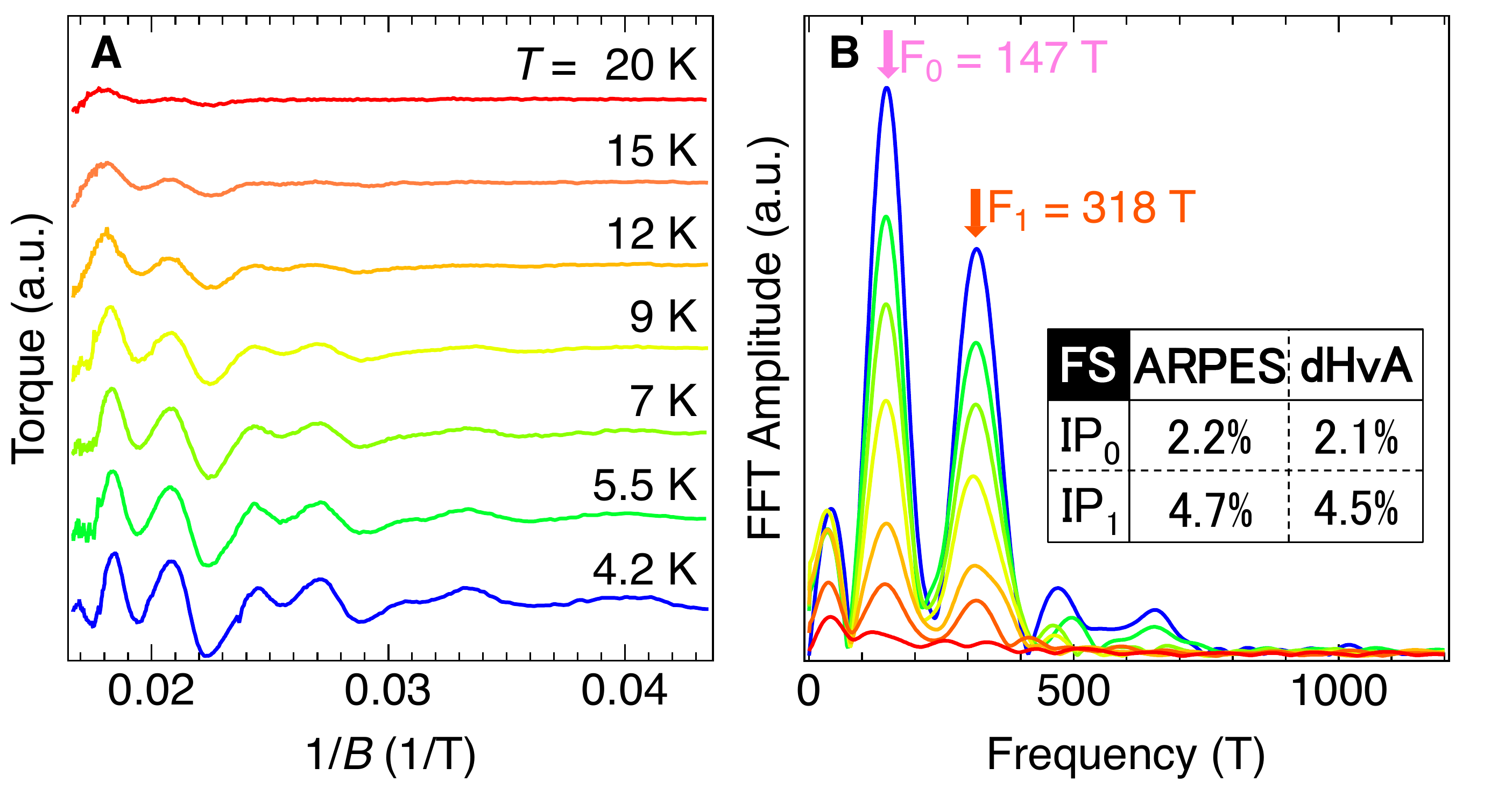}
\end{center}
\label{fig2}
\end{figure}

\noindent {\bf Fig. 2.} 
{\bf Small Fermi pockets revealed by quantum oscillations.} 
({\bf A}) Magnetic torque signals [de Haas-van Alphen effect (dHvA)] at selected different temperatures where smooth backgrounds are subtracted. The angle between the field direction and the crystallographic $c$-axis was set to be 7 degrees during the measurements. 
({\bf B})  The fast Fourier transformation  (FFT) spectra of the observed quantum oscillations in (A).
Arrows mark the two main peaks (F$_0$ and F$_1$). The inset table summarizes the area of the two Fermi pockets, estimated by means of dHvA and ARPES, as a percentage ($\%$) of the Brillouin zone area. Here we note that the small peaks other than F$_0$ and F$_1$ are yielded due to trivial reasons:  While the small peaks at frequencies higher than F$_1$ could represent higher harmonics of F$_0$ and F$_1$, these must be affected by or might be even just artifacts due to noise in the raw data, since the peak positions are not stable for different temperatures. On the other hand, the small peaks below F$_0$, though stable with temperature, are sensitive to the method of background subtraction, and thus these are not intrinsic (see the raw data before background subtraction in the supplemental Fig. S4B); we assumed a polynomial curve as the background for each quantum oscillation spectra. Hence, artificial intensities of a wave-like structure with a low frequency are inevitably left after the treatment of background subtraction. Most importantly, we have confirmed that the frequencies of the two main peaks (F$_0$ and F$_1$) are robust against both the noise in data and the method of background subtraction.

\clearpage
\newpage

\begin{figure} 
\begin{center}
\includegraphics[width=4.5in]{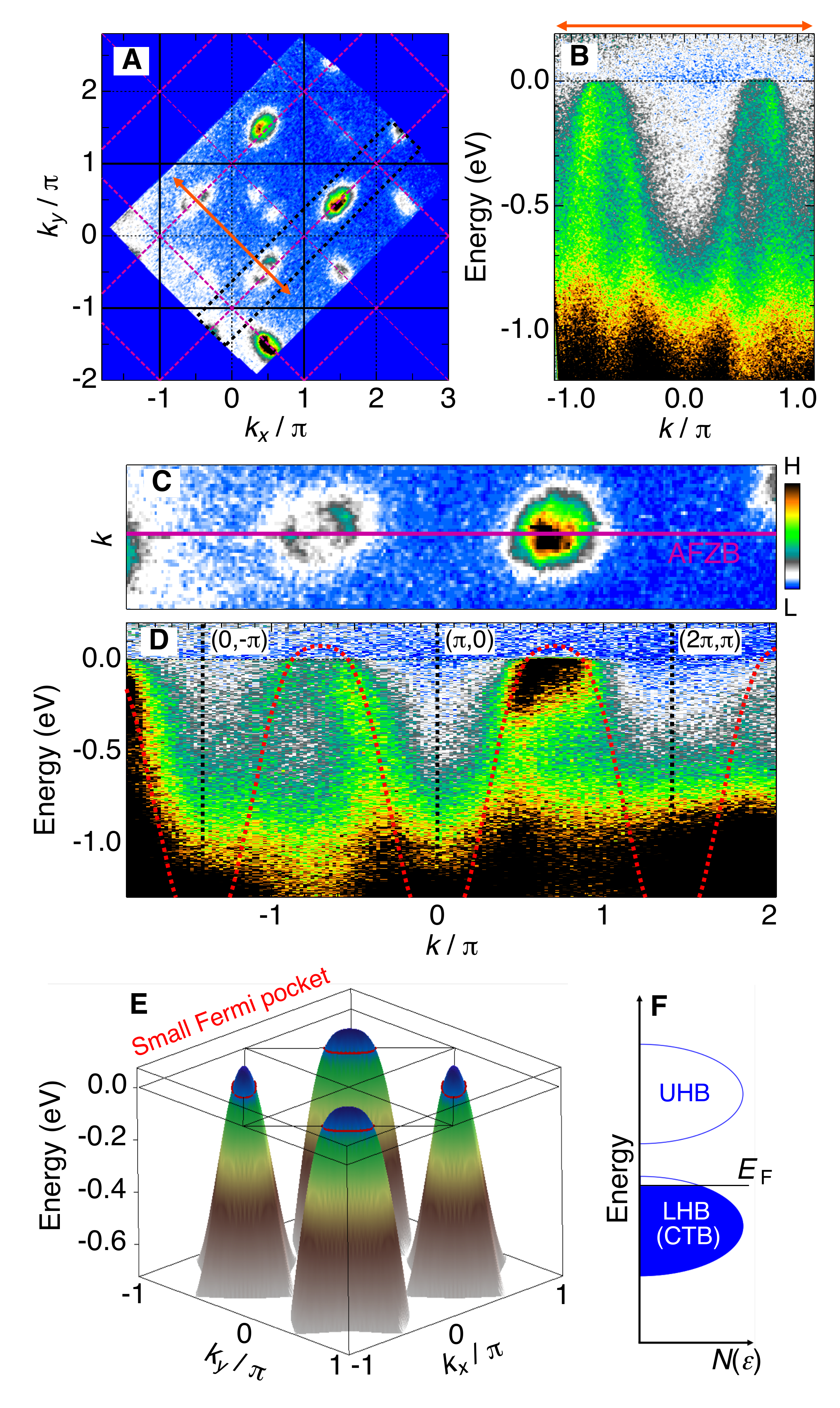}
\end{center}
\label{fig3}
\end{figure}

\clearpage
\newpage

\noindent {\bf Fig. 3.} 
{\bf Overall band structure with Fermi pocket revealed by synchrotron-ARPES.} 
({\bf A}) ARPES intensity map about $E_F$ over multiple Brillouin zones measured at $T=10$ K. The data were taken at $h\nu =$70 eV, which selectively observes Fermi pocket (Supplemental Fig. S7). 
({\bf B}) ARPES dispersion along a diagonal momentum cut (a red arrow in (A)). 
({\bf C}) Zoomed image enclosed by a black dashed square in (A). 
({\bf D}) ARPES dispersion along AFZB indicated in (C). Band dispersion is traced by dashed red curves, which are extracted from the band shape in (E). 
({\bf E}) Band dispersion determined by tight-binding fitting to our ARPES data. 
({\bf F}) Schematic for the density of states in lightly hole-doped  Mott state;  UHB, LHB, and CTB stand for upper Hubbard band, lower Hubbard band, and charge transfer band, respectively. 

\clearpage
\newpage

\begin{figure} 
\begin{center}
\includegraphics[width=6.5in]{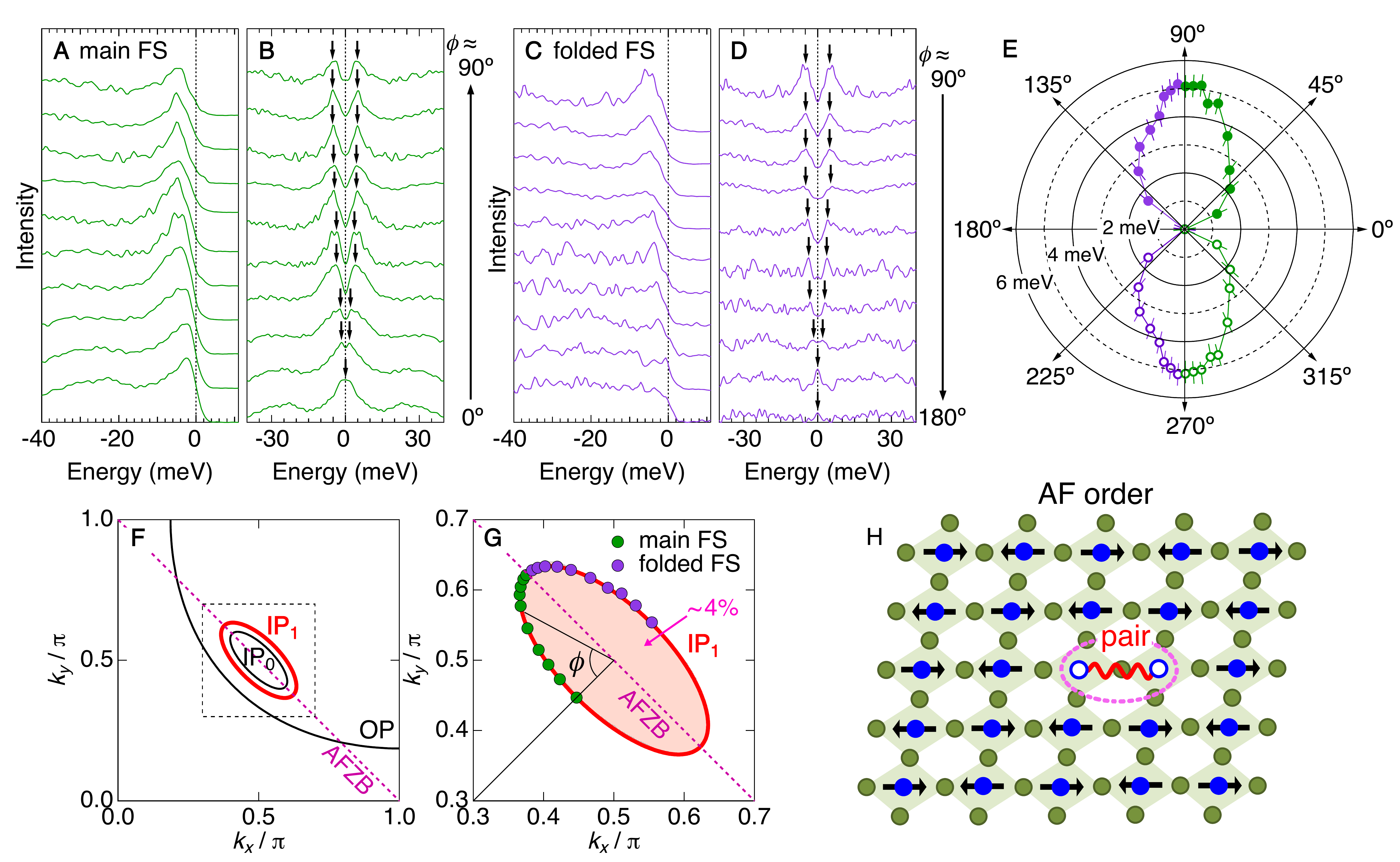}
\end{center}
\label{fig4}
\end{figure}

\noindent {\bf Fig. 4.} 
{\bf Superconducting gap along the Fermi pocket observed by high resolution laser-ARPES.} Energy distribution curves (EDCs) measured far below $T_c$ ($T=$5 K) at $k_F$'s along Fermi pocket for the second inner plane (IP$_1$; see (F)) on the main  side ({\bf A}) and the folded side ({\bf C}) of Brillouin zone: the measured $k_F$'s are marked by green and purple circles in ({\bf G}). 
({\bf B and D}) Symmetrized EDCs of (A) and (C), respectively, which visualize the gap opening.
({\bf E}) The magnitude of the superconducting gaps (arrows in (B) and (D)) plotted in polar coordinates.  The angle $\phi$ is defined in (G). 
({\bf F}) Fermi surfaces (FSs) determined by our ARPES data: the dominant CuO$_2$ plane for each FS is labeled by IP$_0$, IP$_1$, and OP. 
({\bf G}) Zoomed FS for IP$_1$ (a region marked by dashed square in (F)) with the $k_F$ points where high resolution measurements for the gap estimation in (A to D) were performed. 
({\bf H}) Schematic real-space showing the coexistence between the AF-order and electron pairs.

\end{document}


\title{Supplementary Materials for

Observation of small Fermi pockets protected
by clean CuO$_2$ sheets of a high-$T_c$ superconductor}

\author{So~Kunisada}
\affiliation{ISSP, University of Tokyo, Kashiwa, Chiba 277-8581, Japan}

\author{Shunsuke~Isono}
\affiliation{Department of Applied Electronics, Tokyo University of Science, Tokyo 125-8585, Japan}

\author{Yoshimitsu~Kohama}
\affiliation{ISSP, University of Tokyo, Kashiwa, Chiba 277-8581, Japan}
\affiliation{Trans-scale Quantum Science Institute, University of Tokyo, Bunkyo-ku, Tokyo 113-0033, Japan}

\author{Shiro~Sakai}
\affiliation{RIKEN Center for Emergent Matter Science (CEMS), Wako, Saitama 351-0198, Japan}

\author{C\'edric~Bareille}
\affiliation{ISSP, University of Tokyo, Kashiwa, Chiba 277-8581, Japan}

\author{Shunsuke~Sakuragi}
\affiliation{ISSP, University of Tokyo, Kashiwa, Chiba 277-8581, Japan}

\author{Ryo~Noguchi} 
\affiliation{ISSP, University of Tokyo, Kashiwa, Chiba 277-8581, Japan}

\author{Kifu~Kurokawa}
\affiliation{ISSP, University of Tokyo, Kashiwa, Chiba 277-8581, Japan}

\author{Kenta~Kuroda}
\affiliation{ISSP, University of Tokyo, Kashiwa, Chiba 277-8581, Japan}

\author{Yukiaki~Ishida}
\affiliation{ISSP, University of Tokyo, Kashiwa, Chiba 277-8581, Japan}

\author{Shintaro~Adachi}
\affiliation{MANA, National Institute for Materials Science, Tsukuba, Ibaraki 305-0047, Japan}

\author{Ryotaro~Sekine}
\affiliation{Department of Applied Electronics, Tokyo University of Science, Tokyo 125-8585, Japan}
  
\author{Timur~K.~Kim}
\affiliation{Diamond Light Source, Harwell Campus, Didcot, OX11 0DE, United Kingdom}
   
\author{Cephise~Cacho}
 \affiliation{Diamond Light Source, Harwell Campus, Didcot, OX11 0DE, United Kingdom}
 
\author{Shik~Shin} 
\affiliation{ISSP, University of Tokyo, Kashiwa, Chiba 277-8581, Japan}
\affiliation{Office of University Professor, The University of Tokyo, Kashiwa, Chiba 277-8568, Japan}

\author{Takami~Tohyama}
\affiliation{Department of Applied Physics, Tokyo University of Science, Tokyo 125-8585, Japan}
    
\author{Kazuyasu~Tokiwa} 
\affiliation{Department of Applied Electronics, Tokyo University of Science, Tokyo 125-8585, Japan}

\author{Takeshi~Kondo}
\affiliation{ISSP, University of Tokyo, Kashiwa, Chiba 277-8581, Japan}
\affiliation{Trans-scale Quantum Science Institute, University of Tokyo, Bunkyo-ku, Tokyo 113-0033, Japan}

\date{\today}

\maketitle

\newpage

\section {M\MakeLowercase{aterials and methods}}

\begin{figure*}[h]
\begin{center}
\includegraphics[width=6in]{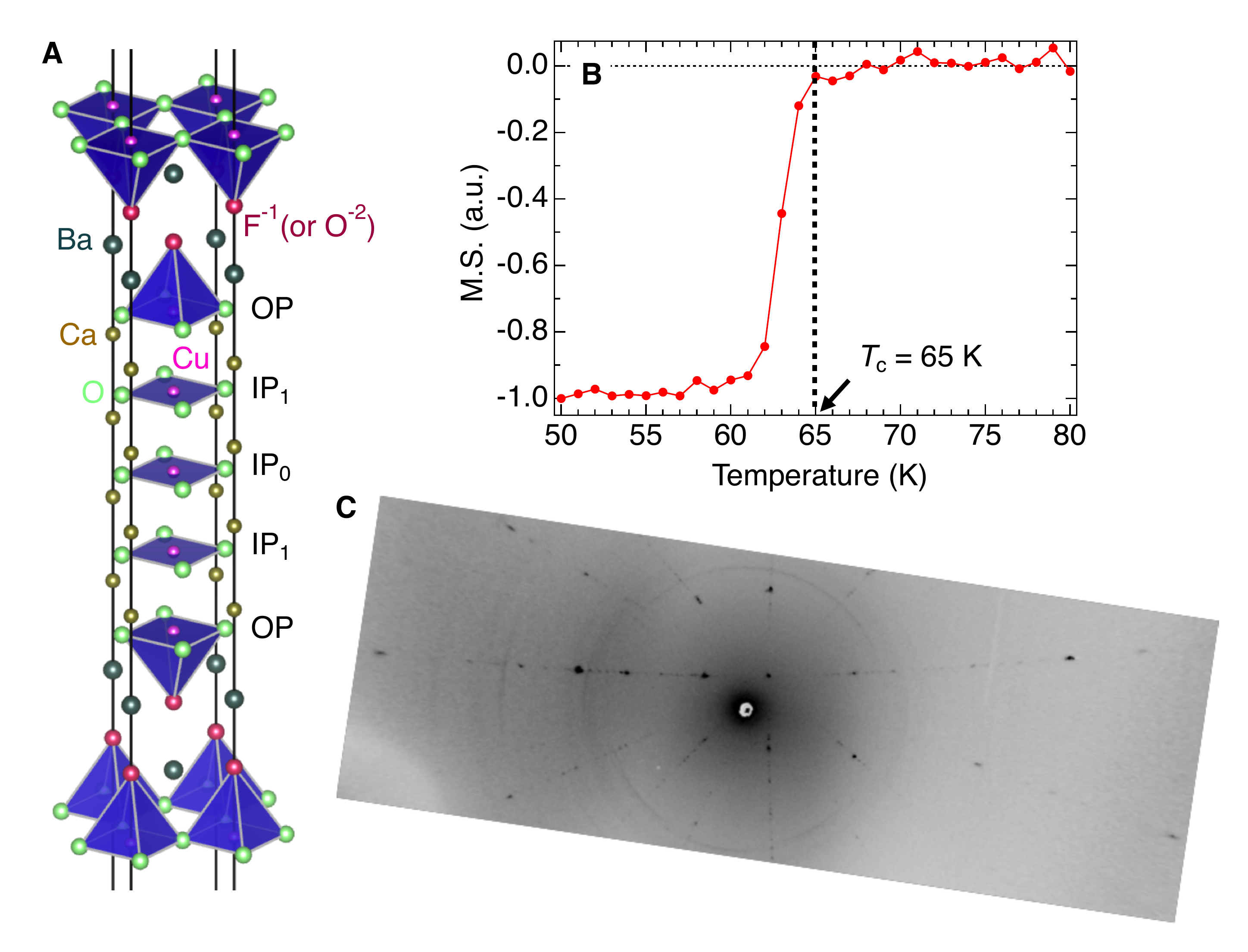}
\caption{{\bf Sample properties.} ({\bf A}) Crystal structure of Ba$_2$Ca$_4$Cu$_5$O$_{10}$(F,O)$_2$. 
({\bf B}) Magnetic susceptibility of the crystals used for ARPES measurements. $T_c$ is estimated to be 65 K from the onset. While signal-to-noise ratio is not so high owing to the small volume in our crystals ($\sim 500 \times 300 \times 50$ $\mu$m in size), the sharp transition with $\sim$3 K in width indicates high quality in the samples. ({\bf C}) Laue image showing a four-fold rotational symmetry of the crystal, and no indication of structural modulations.}
\label{sample}
\end{center}
\end{figure*}

{\bf Samples:} 
Single crystals of underdoped Ba$_2$Ca$_4$Cu$_5$O$_{10}$(F,O)$_2$ (see crystal structure in Fig. \ref{sample}A) with $T_c$=65K were grown at between 1100 $^\circ$C and 1200 $^\circ$C under a pressure of 4.5 GPa without an intentional flux. Magnetic susceptibilities for these crystals (Fig. \ref{sample}B) show sharp superconducting transitions with $\sim$3 K in width, indicative of high quality; signal-to-noise ratio is not so high owing to the small volume in our crystals ($\sim 500 \times 300 \times 40$ $\mu$m in crystal size). Laue image (Fig. \ref{sample}C) shows a four-fold rotational symmetry of the crystal and no indication of structural modulations.\\ 

{\bf ARPES measurements:} 
Laser-based ARPES data were accumulated using a laboratory-based system consisting of a Scienta R4000 electron analyzer and a 6.994 eV laser (the 6th harmonic of Nd:YVO$_4$ quasi-continuous wave). The overall energy resolution in the ARPES experiment was set to 1.4 meV. To get a clean surface, a typical cleavage method was used: a top post glued on the crystal is hit {\it in situ} to obtain a flat surface suitable for the ARPES measurements. ARPES intensity map in Fig. 1A has been normalized to the spectral intensities around -0.15 eV, and the ones in Figs. 1K to 1M have been normalized to the integrated intensities of the reference spectra taken from gold.

Synchrotron-based ARPES measurements were performed at high-resolution branch (HR-ARPES) of the beamline I05 in the Diamond Light Source, equipped with a Scienta R4000 analyzer. The photon energies used were 70 eV and 100 eV. The overall energy resolution was set to $\sim$12 meV in our experiments.\\

{\bf  Quantum oscillation and Hall resistance measurements:}
 Quantum oscillation measurements were performed in pulsed magnetic fields up to 60 T (36 ms pulse duration). For torque magnetometry experiments, a commercial piezoresistive cantilever (SEIKO PRC-120) was used \cite{Torque}. The cantilever directly detects the magnetic torque ($\tau$) as the result of the anisotropic magnetization of the sample, $\tau=${\boldmath $M$}$\times${\boldmath $H$}, and the magnetic quantum oscillation known as the de Haas-van Alphen (dHvA) oscillation was observed. 
Figure 2 in the main paper shows the data after subtracting background, which was obtained by fitting a quadratic function to each curve of the raw data in the range of magnetic field between 32 and 60 T.
 

A tunnel diode oscillator (TDO) was employed for detecting resistive quantum oscillation (the Shubnikov–de Haas effect: SdH) \cite{TDO}. A  single crystal 
was attached to a coil of 8 turns that is part of a TDO circuit. The oscillator resonates at a frequency of 80 MHz in the normal state, jumping by 160 kHz in response to the appearance of the superconductivity in the sample. As the in-plane resistivity changes due to the SdH effect, the resonance frequency was correspondingly shifted. The temperature was controlled between 1.5 K and 100 K by an in-house plastic $^4$He refrigerator. 

Hall resistance of the single crystal 
was measured at fixed temperatures in a pulsed field up to 55 T. The field dependence of Hall resistance was detected using a numerical lock-in technique operating at 100 kHz in a standard Hall bar geometry.
\\

\section {P\MakeLowercase{roperties of} ARPES \MakeLowercase{data justifying intrinsic} F\MakeLowercase{ermi pockets, not artifacts} (F\MakeLowercase{ig.} S2)}

\begin{figure*}[h]
\includegraphics[width=6.5in]{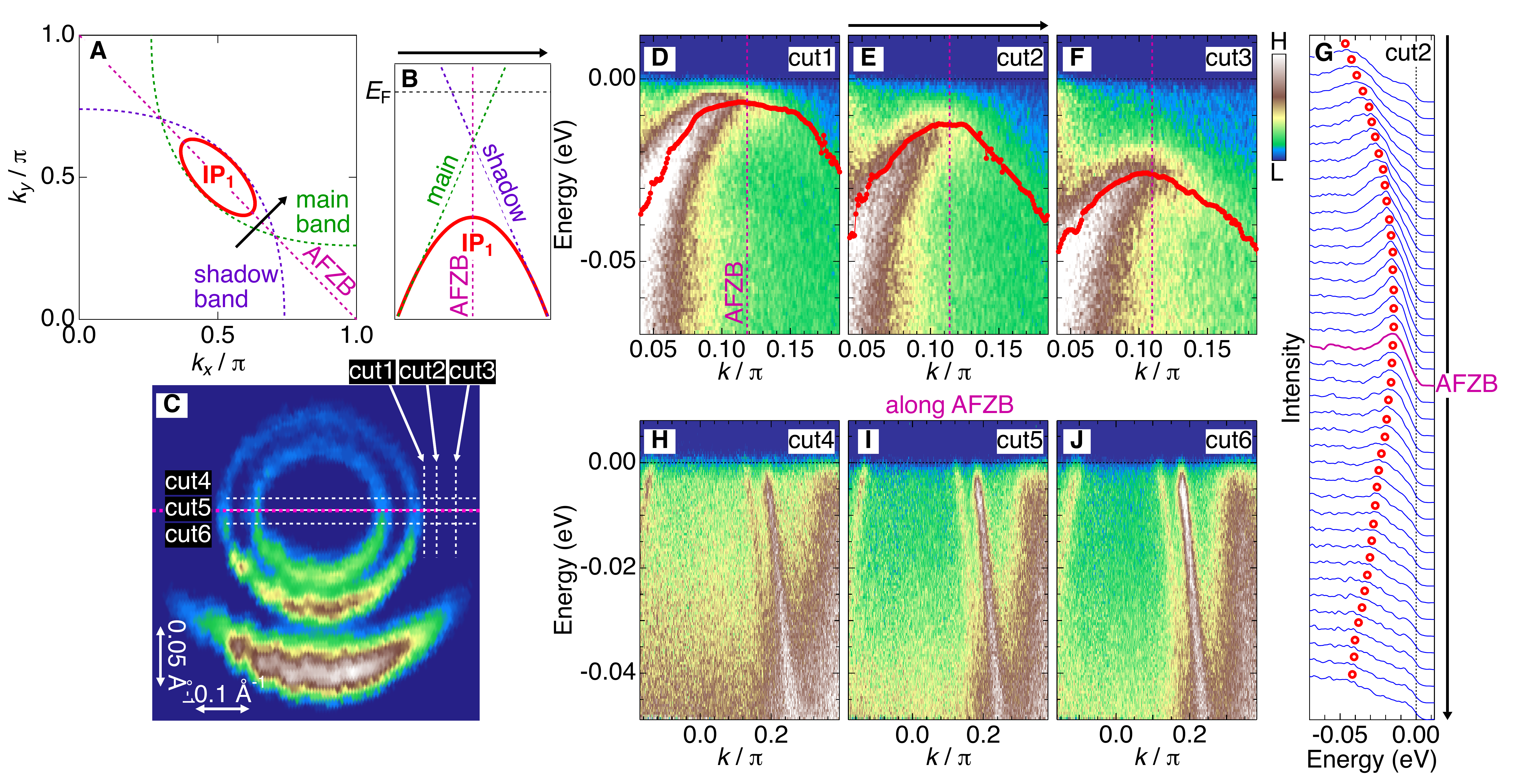}
\caption{{\bf Properties of ARPES data justifying intrinsic Fermi pockets, not artifacts.}  
({\bf A}) Fermi pocket for the second inner layers IP$_1$ determined by ARPES (solid red curve). The dotted curves indicate large Fermi surfaces of the main band (green) and of an artificial shadow band (purple) reflected about AFZB (dotted magenta line). ({\bf B}) Band dispersion for the Fermi pocket (solid red curve) along an arrow in (A). The dotted lines indicate the main band (green) and an artificial shadow band (purple) reflected about AFZB (dotted magenta line). ({\bf C}) Fermi surface mapping by laser-ARPES zoomed around ($\pi/2$,$\pi/2$): the same figure as the main Fig. 1B. ({\bf D to F}) ARPES dispersions along the momentum cuts (cut1, cut2, and cut3, respectively) marked by dashed lines in (C). The red circles trace band dispersions determined from the peak positions of energy distribution curves (EDCs). ({\bf G}) EDCs extracted from (E). The peak positions marked by red circles trace the band dispersion plotted in (E). ({\bf H}) ARPES dispersion along cut4 in (C) for the folded side of Brillouin zone (BZ). ({\bf I}) ARPES dispersion along cut5 in (C), which is just on AFZB. ({\bf J}) ARPES dispersion along cut6 in (C) for the main side of BZ.}
\label{pocket}
\end{figure*}

Here we justify that our ARPES data show intrinsic Fermi pockets, not artifacts due to the overlapping of the main band and an extrinsic shadow band (Figs. \ref{pocket}A and \ref{pocket}B) detected only through the photoemission process. The Fermi pockets for IP$_0$ and IP$_1$ in our ARPES data (Fig. \ref{pocket}C) both perpendicularly cross the antiferromagnetic zone boundary (AFZB) in contrast to the artifacts consisting of overlapped main and shadow bands, which never perpendicularly cross AFZB (see Fig. \ref{pocket}A). Similarly, the band dispersions for the pockets (Figs. \ref{pocket}D, \ref{pocket}E, and \ref{pocket}F) along cut1, cut2, and cut3 in Fig. \ref{pocket}C, determined from the ARPES spectral peaks (see Fig. \ref{pocket}G for cut2),  also perpendicularly cross AFZB; these again disagree with the artifacts, which never perpendicularly cross AFZB as depicted in Fig. \ref{pocket}B.  

These arguments are also supported by the ARPES dispersions observed along AFZB and momentum cuts adjacent in parallel to it; Figures \ref{pocket}H, \ref{pocket}I, and \ref{pocket}J plot the ARPES dispersions measured along the cut4, cut5, and cut6 in Fig. \ref{pocket}C, respectively, which are each located in the folded side, on the border (that is, AFZB), and in the main side of Brillouin zone. Importantly, we observe energy dispersions of spectral peaks with very narrow momentum widths, which are not expected for the overlapped main and shadow bands and instead demonstrate the formation of intrinsic Fermi pockets in  Ba$_2$Ca$_4$Cu$_5$O$_{10}$(F,O)$_2$. 
\\

\section {C\MakeLowercase{omparison of nodal spectra between the current samples and} B\MakeLowercase{i}2212 (F\MakeLowercase{ig.} S3)}

In ARPES studies for cuprates, the spectral analysis has been carried out in detail mostly for the nodal spectra of Bi$_2$Sr$_2$CaCu$_2$O$_{8+\delta}$ (Bi2212), which is easy to cleave, thus provides a flat surface required to obtain intrinsic line shapes in the ARPES spectra. Especially, the laser-ARPES is acknowledged to be best suited for such studies owing to high energy and momentum  resolutions. It would be, therefore, worth comparing the nodal spectra of Ba$_2$Ca$_4$Cu$_5$O$_{10}$(F,O)$_2$ and Bi2212 both measured by laser-ARPES at the same experimental setting, in order to understand the electronic properties of the former (the five-layered compound).

\begin{figure*}[h]
\includegraphics[width=6.5in]{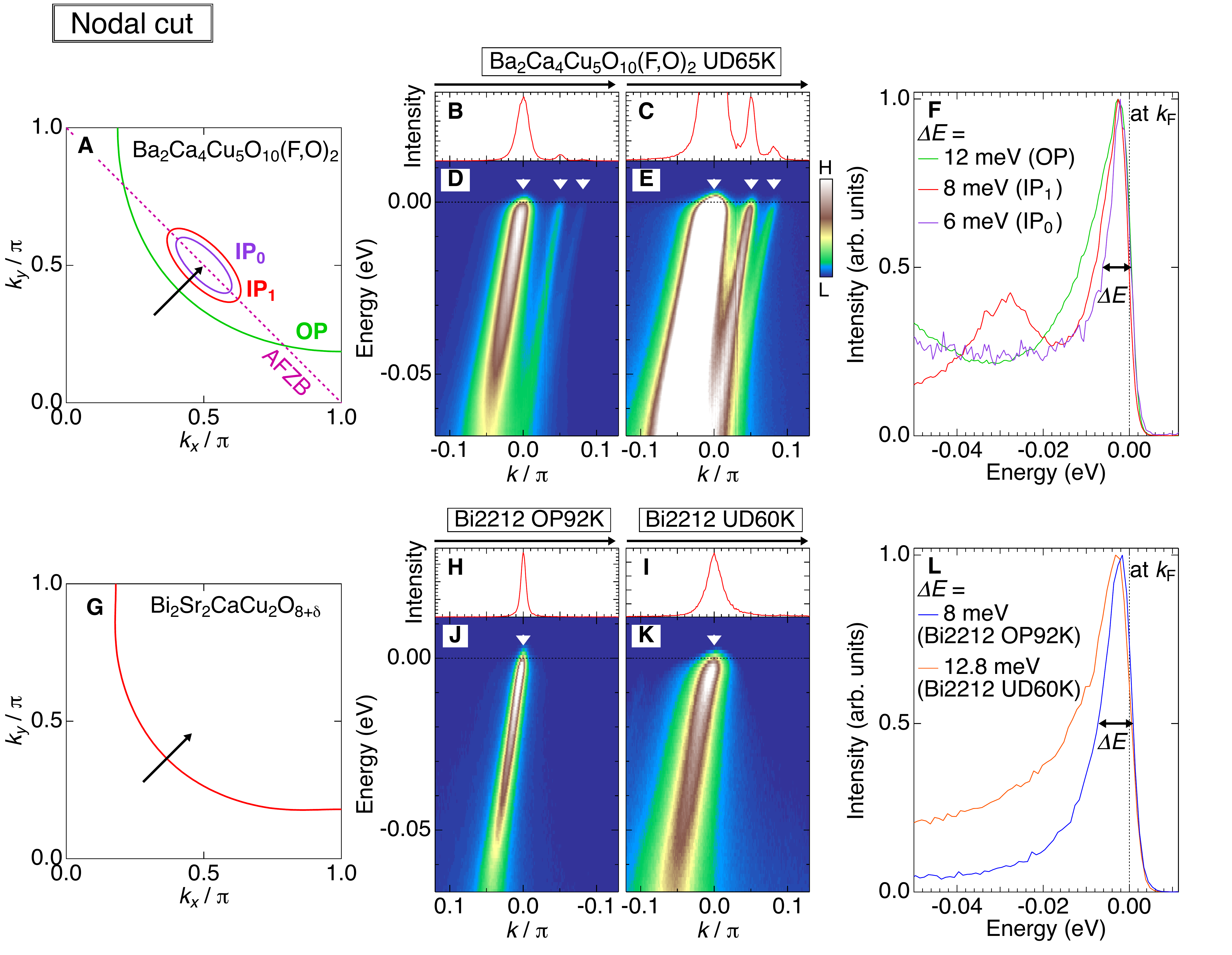}
\caption{{\bf Comparison of nodal spectra between the current samples and Bi2212.}
({\bf A}) Fermi surfaces in Ba$_2$Ca$_4$Cu$_5$O$_{10}$(F,O)$_2$. 
({\bf B and C}) Momentum distribution curves (MDCs) at $E_F$ for ARPES images in (D) and (E), respectively.  
({\bf D}) ARPES dispersion along the nodal momentum cut (arrow in (A)). 
({\bf E}) The same image as {\bf D}, but the color scale is changed to clearly show the dispersions for Fermi pockets (IP$_0$ and IP$_1$ in (A)). 
({\bf F}) Energy distribution curves (EDCs) at the nodal $k_F$ points marked by arrows in (D or E) for the Fermi arc (OP), the larger pocket (IP$_1$), and the smaller pocket (IP$_0$). The energy widths of these three EDCs ($\Delta E$s) are listed inside the panel; as an example, $\Delta E$ for IP$_0$ is indicated by an arrow. 
({\bf G}) Fermi surface in Bi$_2$Sr$_2$CaCu$_2$O$_{8+\delta}$ (Bi2212). 
({\bf H and I}) MDCs at $E_F$ for ARPES image in (J) and (K), respectively.  
({\bf J and K}) ARPES dispersions along the nodal momentum cut (arrow in (G)) for optimally doped Bi2212 with $T_c$=92 K (Bi2212 OP92K) and for underdoped Bi2212 with $T_c$=60 K (Bi2212 UD60K), respectively.
({\bf L}) EDCs at the nodal $k_F$ points for Bi2212 OP92K and UD60K marked by arrows in (J) and (K), respectively. The energy widths of these two EDCs ($\Delta E$s) are listed inside the panel; as an example, $\Delta E$ for Bi2212 OP92K is indicated by an arrow. All the data of Ba$_2$Ca$_4$Cu$_5$O$_{10}$(F,O)$_2$ and  Bi2212 presented here were measured at $T=$5 K by laser-ARPES.
}
\label{compare}
\end{figure*}

Figures \ref{compare}H to \ref{compare}L exhibit the typical laser-ARPES data we observed along the nodal direction in Bi2212 with two different doping levels \cite{Kondo_PRL,Kondo_NC}. The optimally doped Bi2212 with $T_c$ of 92K (OP92K) has a very sharp energy distribution curve (EDC) with 8meV in width ($\Delta E$) at $k_F$ (Fig. \ref{compare}L), indicative of a long lifetime in the energy state. The spectrum gets broader toward underdoping due to a stronger electron scattering (Fig. \ref{compare}L); $\Delta E$=12.8 meV, larger than that of OP92K, is obtained for the underdoped Bi2212 with $T_c$ of 60K (UD60K). 

Very interestingly, we found the opposite trend in the carrier concentration ($p$) dependence of the spectral shape in  Ba$_2$Ca$_4$Cu$_5$O$_{10}$(F,O)$_2$ (Figs. \ref{compare}A to \ref{compare}F), in which $p$ gets smaller from the outer plane (OP) to the inner plane (IP$_1$), and further down to the innermost plane (IP$_0$). Against expectation from the data of Bi2212, the nodal EDCs gets sharper with decreasing $p$ from Fermi arc (OP in Fig. \ref{compare}A), to the larger Fermi pocket (IP$_1$ in Fig. \ref{compare}A), and further down to the smaller Fermi pocket (IP$_0$ in Fig. \ref{compare}A); note that the $\Delta E$ value is smallest in IP$_0$, which is most underdoped.
 
Furthermore, we found that the EDC for IP$_0$ ($\Delta E$=6meV) is even sharper than that of Bi2212 OP92K ($\Delta E$=8meV). This is striking, considering that the amount of hole-doping in IP$_0$ is very small ($p\sim$0.04). These data indicate that the protected inner CuO$_2$ layer is exceptionally very clean, thus capable of yielding long-lived quasiparticles, despite it is extremely underdoped. 
\\

\section {Q\MakeLowercase{uantum oscillations in} S\MakeLowercase{hubnikov-de} H\MakeLowercase{aas effect}  \MakeLowercase{and de} H\MakeLowercase{aas-van} A\MakeLowercase{lphen effect}  \newline (F\MakeLowercase{ig.} S4)}

\begin{figure*}[h]
\includegraphics[width=4in]{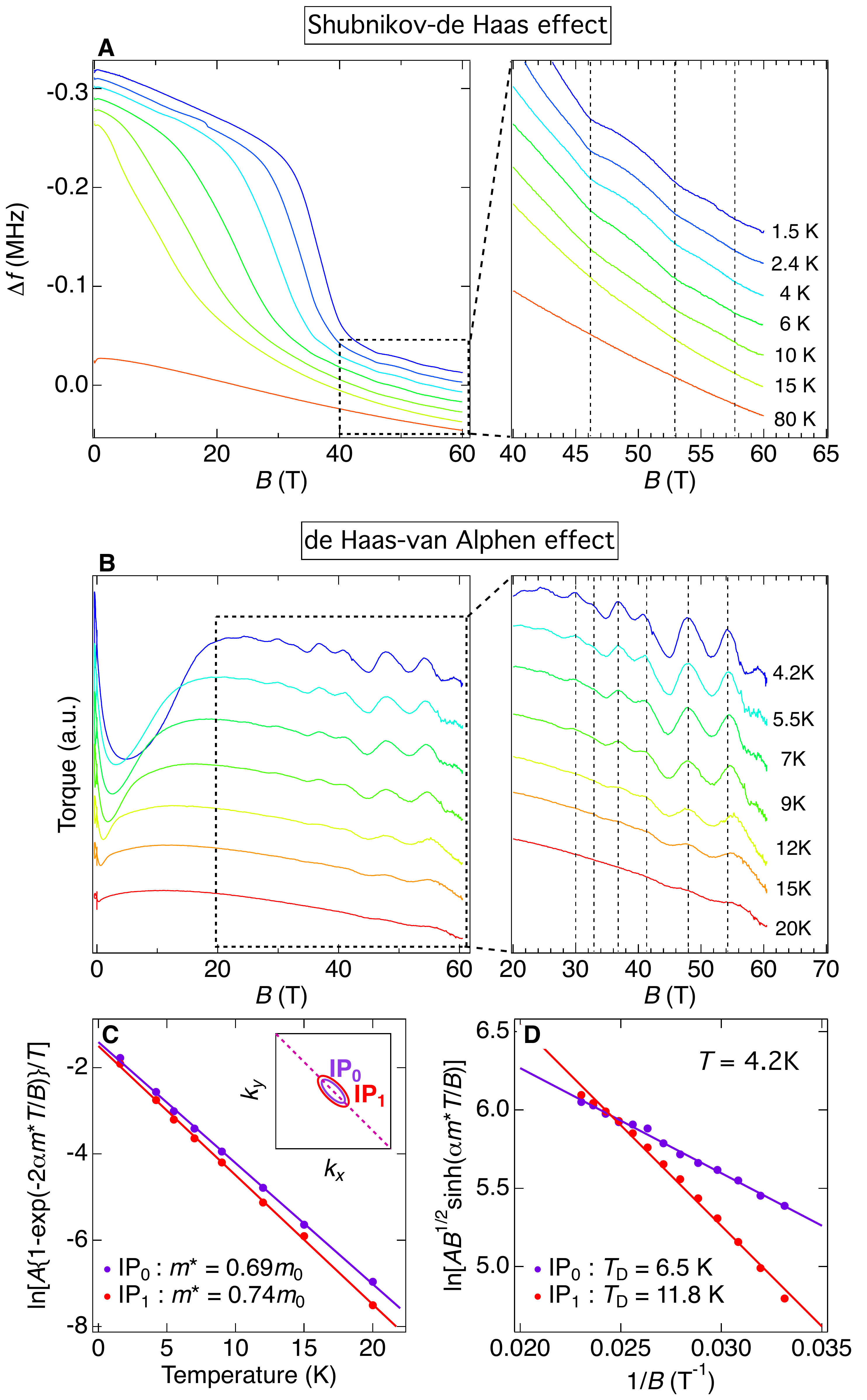}
\caption{{\bf Quantum oscillations in  Shubnikov-de Haas effect and de Haas-van Alphen effect.}
({\bf A}) Resonance frequency of a tunnel diode oscillator (TDO), which can be regarded as proportional to the variation of the in-plane resistivity of the sample. The oscillation is due to the Shubnikov-de Haas effect. The magnetic field was applied along the $c$-axis of the crystal. In this plot, the vertical axis corresponds to the frequency shift ($\Delta f$). The broad decrease in $\Delta f$ with $B$ is caused by the suppression of the superconductivity and relates to the upper critical field. The right panel shows a zoom of the dashed area in the left panel. ({\bf B}) Magnetic torque signals (de Haas-van Alphen effect: dHvA) at selected temperatures in high magnetic fields up to 60 T. The angle between the magnetic field and the crystallographic $c$-axis was set to be 7 degrees. The broad dip at a low field is attributable to the irreversibility field. The right panel is a zoom of the dashed area in the left panel. ({\bf C}) Temperature dependence of the quantum oscillation amplitude in dHvA.
The fit to the standard Lifshitz-Kosevich formula yields an effective mass ($m^*$) of 0.69 $m_0$ for IP$_0$ and 0.74 $m_0$ for IP$_1$, where $m_0$ is the free electron mass \cite{Akiba}. The inset shows the Fermi pockets for IP$_0$ and IP$_1$ determined by ARPES. ({\bf D}) Dingle plot of the quantum oscillation amplitude in dHvA as a function of 1/$B$. From the fit to data at $T=$4.2 K, we obtain the Dingle temperature ($T_D$) of  6.5 K and 11.8 K for IP$_0$ and IP$_1$, respectively.}
\label{quantum}
\end{figure*}

In the main paper, we show quantum oscillations only of de Haas-van Alphen effect (dHvA), detected in magnetic torque signals. Here we also present quantum oscillations of Shubnikov-de Haas effect (SdH) in Fig. \ref{quantum}A, detected in the resonance frequency of a tunnel diode oscillator (TDO); while the oscillation against the magnetic field ($B$) is rather weak, it is clearly visible above $B=40$T (see zoomed data shown in the right panel of Fig. \ref{quantum}A). For comparison, we also plot, in Fig. \ref{quantum}B, quantum oscillations of dHvA, which are the data same as in the main Fig. 2 but before background subtraction; the background was obtained by fitting a quadratic function to each curve of the raw data between 32 and 60 T.

From the behavior of quantum oscillations, we can extract two physical quantities: effective mass ($m^*$) and Dingle temperature ($T_D$). Here we estimate these values especially from the data of dHvA (Fig. \ref{quantum}B and the main Fig. 2), exhibiting clearer oscillations than those of SdH. Figure \ref{quantum}C plots the temperature dependence of the amplitude of the dHvA oscillation; the fit to the standard Lifshitz-Kosevich formula yields an effective mass ($m^*$) of 0.69$m_0$ for IP$_0$ and 0.74$m_0$ for IP$_1$ \cite{Akiba} ($m_0$: the free electron mass). We have confirmed that these are consistent with the effective masses near $E_F$ in the band dispersions with Fermi pockets determined by ARPES. The value of $T_D$ is estimated from the slope of Dingle plot against 1/$B$ (Fig. \ref{quantum}D); by the fit to the data at $T=$4.2 K, we obtain $T_D$ of  6.5 K and 11.8 K for IP$_0$ and IP$_1$, respectively. Since the $T_D$ value is proportional to the scattering rate of electrons, the smaller $T_D$ estimated for IP$_0$ agrees with the argument that the innermost CuO$_2$ layer is cleanest and has the highest mobility in the five-layered system. 
\\

\section {T\MakeLowercase{wo-dimensional structure of} F\MakeLowercase{ermi pockets revealed \newline  by  angle-dependent quantum oscillation}  (F\MakeLowercase{ig.} S5)}

The dimensionality of the Fermi surface can be examined by angle-dependent measurements of quantum oscillations. Here we reveal that the Fermi pockets in our samples are almost perfectly two-dimensional, which thus strengthens the argument in the main paper that the sizes of Fermi pockets agree between quantum oscillations and ARPES measurements. 

\begin{figure*}[h]
\includegraphics[width=6.5in]{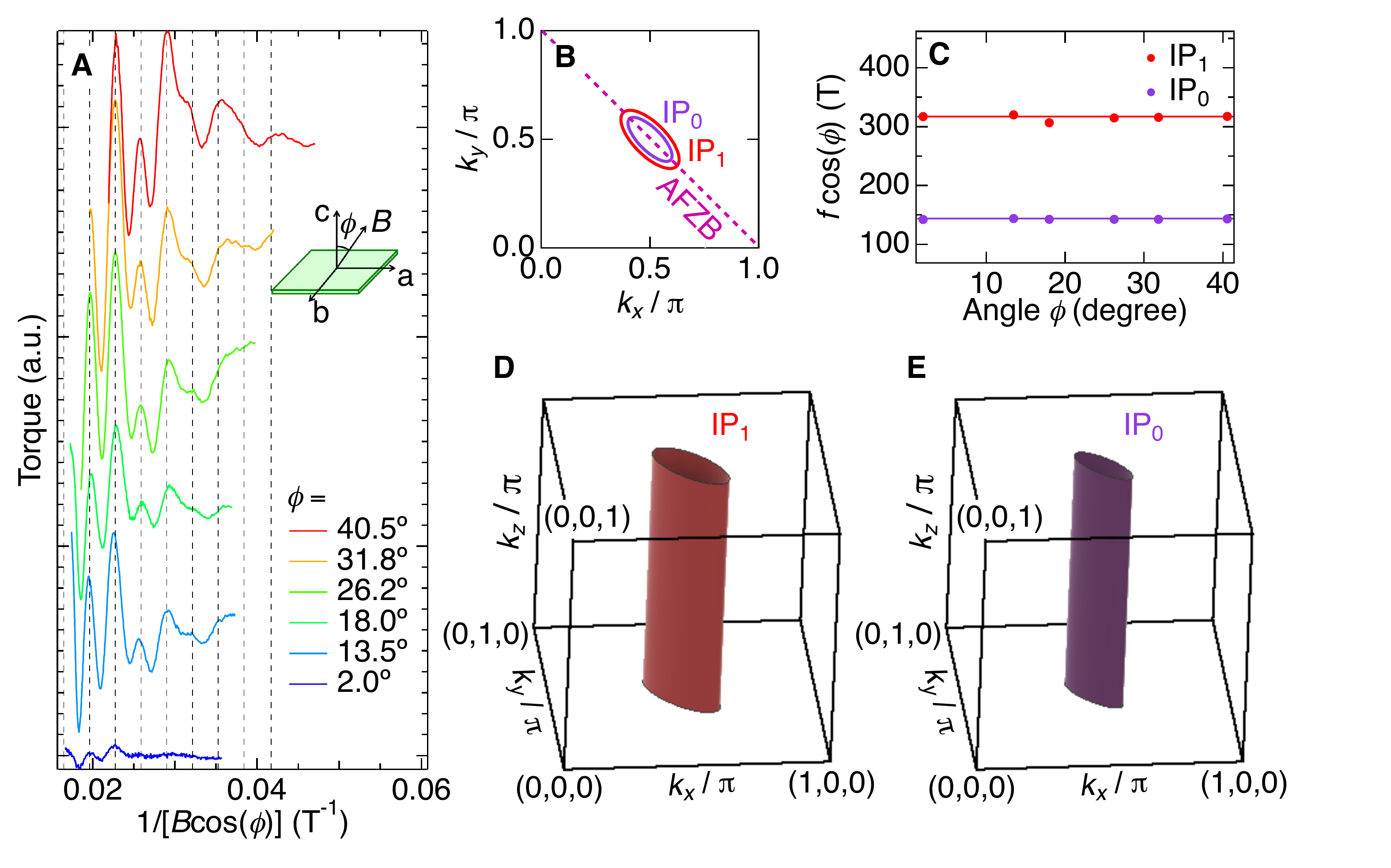}
\caption{{\bf Two-dimensional structure of Fermi pockets revealed by angle-dependent quantum oscillations.}
({\bf A}) Magnetic torque measurements of dHvA at several angles ($\phi$s) between external pulsed-magnetic-field and the $c$-axis of the crystal  ($\phi$ is defined in the inset).  The data were measured at $T=$10 K. ({\bf B}) Fermi pockets for IP$_0$ and IP$_1$ determined by ARPES. ({\bf C}) Angle dependence of quantum oscillation frequency ($f$) multiplied by cos($\phi$) for the two pockets. The constant $f$cos($\phi$) with $\phi$  demonstrates that 
both pockets have two-dimensional structures. ({\bf D and E}) Two dimensional Fermi pockets for IP$_1$ and IP$_0$, respectively, revealed by the angle-dependent magnetic torque measurements.}
\label{2D}
\end{figure*}

In Fig. \ref{2D}A,  we present  dHvA oscillations of  magnetic torque measured at
several angles ($\phi$s) between external pulsed-magnetic-field and the $c$-axis of the crystal; 
 the data are plotted against $1/[B\cos (\phi )]$ for each measured $\phi$, together with dashed lines added at regular intervals for a guide to eyes. We find that the antinodes of these oscillations coincide at the same $1/[B\cos (\phi )]$ values, indicating that the Fermi pockets are two-dimensional. To confirm this further, we extract two components of quantum oscillation frequency ($f$) corresponding to the two pockets (IP$_1$ and IP$_0$ in Fig. \ref{2D}B), and plot these values multiplied by cos($\phi$) in Fig. \ref{2D}C as a function of each measured angle $\phi$. These plots are both constant for different $\phi$'s, indicating that these pockets are almost perfectly two-dimensional as depicted in Figs. \ref{2D}D and \ref{2D}E for IP$_1$ and IP$_0$, respectively.
\\

\section {P\MakeLowercase{ositive charges of} F\MakeLowercase{ermi pocket unchanged even at the high magnetic field, \newline revealed by} H\MakeLowercase{all effect} (F\MakeLowercase{ig.} S6)} 

\begin{figure*}[h]
\includegraphics[width=4in]{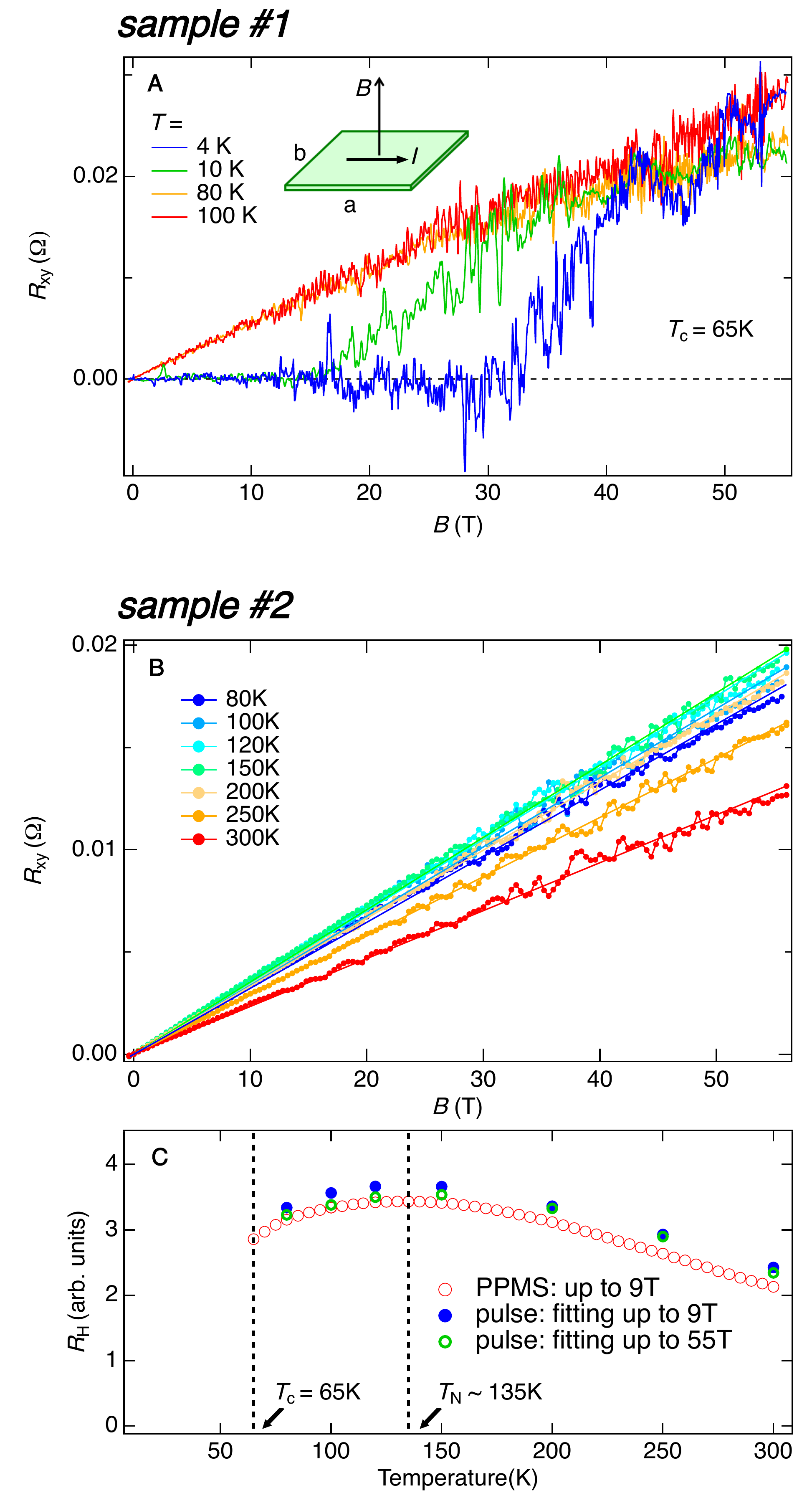}
\caption{{\bf Positive charges of Fermi pocket unchanged even at the high magnetic field, revealed by Hall effect.}
({\bf A})  Hall resistance (${R_{xy}}$) against the pulsed-magnetic field $B$ up to 55 T, measured at several temperatures from below to above $T_c$ (=65 K) for a sample piece (sample $\# 1$).  ({\bf B})  ${R_{xy}}(B)$ measured  at many different temperatures above $T_c$ for another sample piece (sample $\# 2$).  The ${R_{xy}}(B)$ at each temperature is fitted by a straight line.
({\bf C}) The temperature dependence of Hall coefficient, $R_H(T)$, estimated from the slopes of the $R_{xy}$(B) plots in (B) and by a standard PPMS equipment for the same sample piece: sample $\# 2$. For the data with pulsed magnetic fields, we have estimated the values of $R_H$ with two different ways: one is by fitting $R_{xy}$(B) to a straight line up to $B=9$ T (blue circles), which is the same field as that applied in the PPMS measurements (red circles), and the other is by the fitting up to the highest field of $B=55$ T (green circles).
}
\label{hall}
\end{figure*}

The sign of charges forming Fermi pockets cannot be identified from quantum oscillation measurements.  Therefore, before concluding that there is a consistency between the ARPES and quantum oscillation data, we should confirm that the positive sign of charges in the pockets observed by ARPES is unchanged at the high magnetic field applied in quantum oscillation measurements. This argument is crucial since it has been reported for other compounds that charge density wave emerges under the high magnetic field which suppresses the superconductivity, and it reconstructs the Fermi surface from the inherent hole-type to an electron-type \cite{YBCO_Negative}. To address this issue, we have measured Hall resistance (${R_{xy}}$) up to a high magnetic field used for quantum oscillation measurements, as plotted in Fig. \ref{hall}A.

The characteristic behavior of ${R_{xy}}(B)$ seen at 10K and 4K (below $T_c$) is coming from the transition from the superconducting to normal states induced by the magnetic field. The value of ${R_{xy}}$ is zero in the superconducting state, and becomes finite when non-superconducting vortex cores flow in bulk ($\sim15$ T  at 10K and $\sim30$ T at 4 K). With further increasing the magnetic field, ${R_{xy}}$ increases and eventually reaches that of the normal state above $T_c$ at the upper critical field ($\sim30$ T  at 10K and  $\sim40$ T at 4 K), which completely suppresses the superconductivity in bulk. ${R_{xy}}(B)$ at the lowest temperature (4K) shows some overall oscillation above $B=40$ T other than spiky noises. This is possibly due to quantum oscillation, while further experiments would be necessary to conclude it since the noise level is relatively high in our data.

In general, care must be taken when interpreting the Hall effect of multiband systems; ${R_{xy}}(B)$ usually becomes non-linear (or curved) against $B$ in the systems with two types of carriers (hole and electron). For example, ${R_{xy}}(B)$ of electron-doped cuprates \cite{Hall_Edoped},  which are known to have both the hole- and electron-types of carriers, are remarkably curved with $B$. 
In contrast to these, our data of the five-layered systems in the normal state seem rather simple, showing an almost perfectly $B$-linear behavior with a positive slope all the way up to the highest magnetic field ($B=55$ T). This indicates that the carriers in all the five CuO$_2$ planes of Ba$_2$Ca$_4$Cu$_5$O$_{10}$(F,O)$_2$ are hole-type. To confirm it further, we have measured another piece of  the sample at many temperatures above $T_c$, as plotted in Fig. \ref{hall}B; almost perfectly $B$-linear behaviors are obtained for all the plots of ${R_{xy}}(B)$, validating the above argument.

Moreover, the $R_{xy}$(B) below $T_c$ is confirmed to have a positive slope as a whole and reach the same value as that just above $T_c$ at high magnetic fields which totally suppress superconductivity (see Fig. \ref{hall}A). This indicates that the carrier type below $T_c$ is not changed by the magnetic suppression of superconductivity and it should be the hole same as in the metallic state above $T_c$. Therefore, the Fermi pockets detected by quantum oscillation measurements should be the same as the ones observed by ARPES at zero magnetic field, not FSs reconstructed at high magnetic fields as reported for other hole-doped cuprates \cite{YBCO_Negative}.

To examine the evolution of Fermi surface with temperature,  we estimate, in Fig. \ref{hall}C,  the temperature dependence of Hall coefficient, $R_H(T)$, from the slopes of the $R_{xy}$(B) plots in Fig. \ref{hall}B measured with pulsed magnetic fields. Here the obtained data are represented in arbitrary units since the sample thickness required to determine the absolute values has some uncertainty. For comparison, $R_H(T)$ has also been  measured by a standard PPMS equipment for exactly the same sample piece we just used for the pulsed magnetic field measurements, and it is overlapped in Fig. \ref{hall}C. For the data with pulsed magnetic fields, we have estimated the values of $R_H$ with two different ways: one is by fitting $R_{xy}$(B) to a straight line  up to only $B=9$ T (blue circles in Fig. \ref{hall}C), which is the same magnetic field as applied in the PPMS measurements, and the other is up to the highest magnetic field of $B=55$ T (green circles in Fig. \ref{hall}C). We could get a consistency among the three including the PPMS result (red circles in Fig. \ref{hall}C). These data have two significant implications. Firstly, the consistency between the $R_H$ data estimated by fitting $R_{xy}$(B) up to 9 T and 55 T confirms that $R_{xy}$(B) is indeed almost perfectly $B$-linear up to a high magnetic field. Secondly, the $R_H(T)$ exhibits no notable variation across $T_N$ ($\sim135$ K), indicating that the Fermi pockets in the carrier-doped Mott band persist above $T_N$; this agrees to the previous ARPES results of Mott-insulating CCOC ($T_N$=245 K) and SCOC ($T_N$=256 K) that the band folding has been clearly observed at room temperature \cite{CCOC_300K} or even 100 K higher than $T_N$ \cite{Wells_half}.
\\

\section {S\MakeLowercase{elective observation of the} F\MakeLowercase{ermi arc and pocket by synchrotron}-ARPES (F\MakeLowercase{ig.} S7)}

The ARPES using a low energy laser (laser-ARPES) has huge advantages in terms of high energy and momentum resolutions over that using synchrotron (synchrotron-ARPES). However, there is also a drawback in laser-ARPES: the observable momentum space is limited, not being able to cover the whole Brillouin zone (BZ), owing to the low energies of excited photoelectrons. To measure the whole band structure, therefore, synchrotron-ARPES with higher energy photons is required, although the resolutions are sacrificed. Another advantage in synchrotron-ARPES is a tunability of photon energy ($h\nu$), which arrows one to utilize the matrix element effect in photoemission to selectively observe a particular band structure from complex multi-bands. In the main paper, we present the data at $h\nu$=70 eV (the main Fig. 3), which not only selectively detect the bands forming Fermi pockets, but also especially enhance the ARPES intensities in the folded side of BZ. 

\begin{figure*}[h]
\includegraphics[width=4.5in]{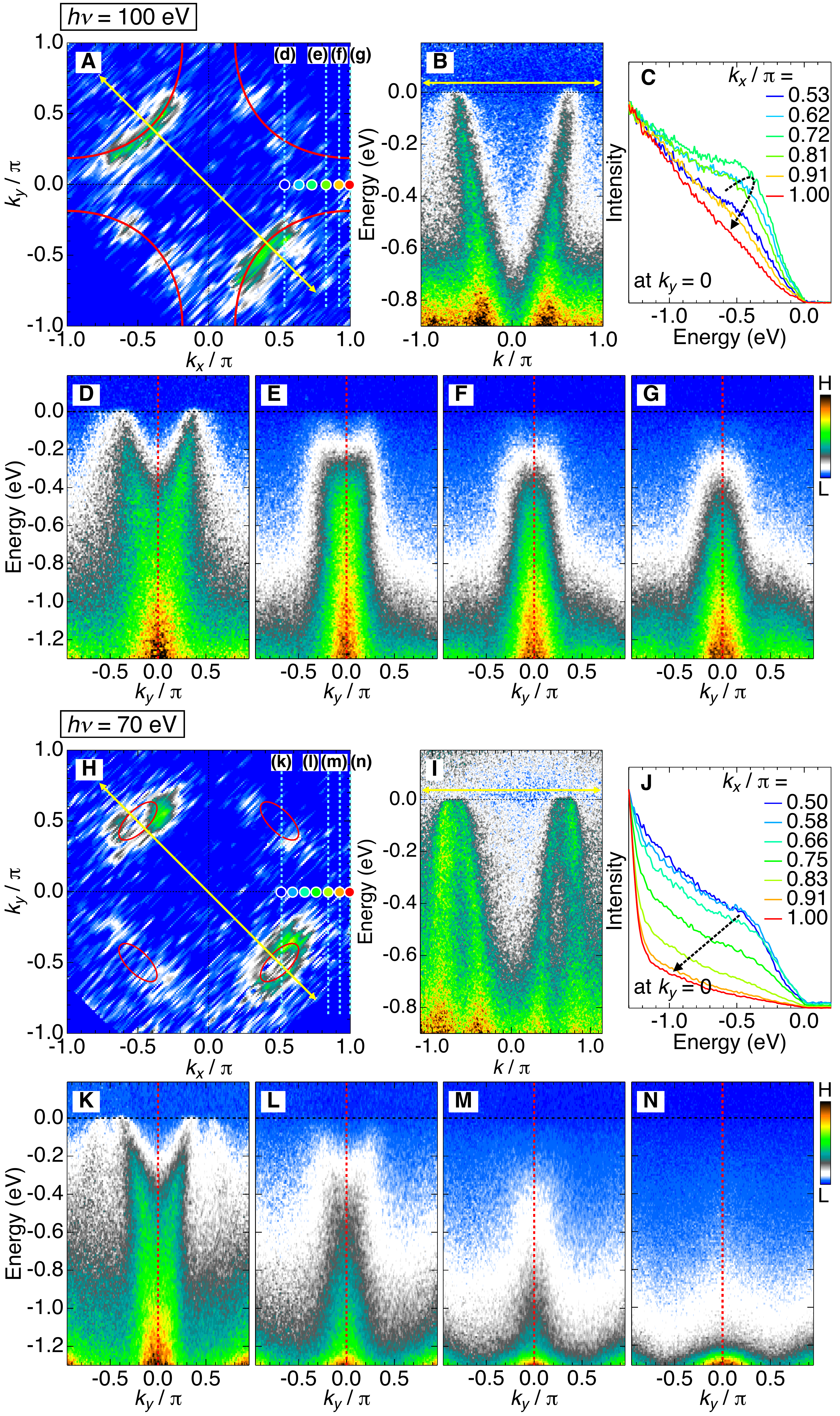}
\caption{{\bf Selective observation of the Fermi arc and pocket by synchrotron-ARPES.} 
({\bf A}) Fermi surface mapping at $h\nu$ = $100$ eV, which selectively observes the Fermi arc. 
({\bf B}) ARPES dispersion along the diagonal momentum cut (arrow in (A)). 
({\bf C})  Energy distribution curves (EDCs) at several $k_x$'s along the (0,0)-($\pi,0$) line (colored circles in (A)). 
({\bf D to G}) ARPES dispersions along several momentum cuts parallel to the zone edge (dashed lines in (A)). 
({\bf H to N}) The same data as (A to G), but measured at $h\nu$ = $70$ eV, which selectively observes the Fermi pocket. The dashed arrows in (C) and (J) trace the behaviors of spectra with approaching the ($\pi,0$) point in the bands forming the Fermi arc and pocket, respectively.  All the data presented here were measured at $T=$10 K. 
}
\label{hv}
\end{figure*}

Here we also present the ARPES data at $h\nu$=100 eV (Figs. \ref{hv}A to \ref{hv}G), which selectively detect the band forming the Fermi arc, and compare them with those at $h\nu$=70 eV (Figs. \ref{hv}H to \ref{hv}N). The ARPES data at $h\nu$=100 eV along the nodal cut (Fig. \ref{hv}B) shows a large parabolic dispersion with a convex downward. This, though typical for cuprates, is  clearly different from the data at $h\nu$=70 eV (Fig. \ref{hv}I), which exhibits two parabolic dispersions with a convex upward centered at ($-\pi/2,\pi/2$) and ($\pi/2,-\pi/2$).  Furthermore, the band dispersions at $h\nu$=100 eV (Figs. \ref{hv}D to \ref{hv}G) extracted parallel to the zone boundaries (dashed lines in Fig. \ref{hv}A) show the opening of pseudogap typical for the underdoped cuprates. At $h\nu$=70 eV, the contrasting features specific to the band with hole pockets are observed along similar momentum cuts (dashed lines in Fig. \ref{hv}H): with approaching the zone edge, the band disperses toward higher binding energies very rapidly in a monotonic fashion. This is more clearly demonstrated in Fig. \ref{hv}J by extracting  EDCs at several $k_F$ points of $k_y$=0 (colored circles in Fig. \ref{hv}H); while spectral peaks are not clearly observed, significant spectral loss spreads from low to high binding energies toward the zone edge (or $k_x$=$\pi$). This behavior contrasts with that for the band forming Fermi arc plotted in Fig. \ref{hv}C; the spectral peak (or shoulder-like structure) first shifts toward lower binding energies before eventually coming back to higher binding energies with approaching the zone edge, just as expected for a band of typical underdoped cuprates, which have a saddle point and open a pseudogap around ($\pi$,0). 

Here we argue that the characteristic band shape with Fermi pockets, which lacks low lying energy states around ($\pi$,0), is intrinsic and robust against the matrix element effect in photoemission, because of the following experimental results: 

(1) The matrix element effect is sensitive to the polarization of incident light. The data at 70 eV in the main Fig. 3D and Figs. S7K-S7N were measured with linearly polarized light which was 45 degrees different in the polarization direction from each other. Importantly, these two consistently exhibit the same band structure as in the main Fig. 3E, which does not have low lying energy states around ($\pi$,0), and instead, disperses down to more than 1eV in binding energy toward ($\pi$,0).

(2) The matrix element effect should also significantly vary in different measurement regions from the 1st to 2nd Brillouin zone. Importantly, our data in the main Fig. 3D consistently exhibit no states at low binding energies both at ($\pi$,0) and (2$\pi$,0), but instead show a periodic dispersion reaching high binding energies more than 1 eV at these $k$ points (see red curves plotted in the main Fig. 3D).

These results of (1) and (2), supporting the robustness against the matrix element effect, justify that the band structure determined in Fig. 3E is intrinsic for the inner planes forming Fermi pockets. 
\\

\section {M\MakeLowercase{odel calculations demonstrating the superconductivity intrinsic to \newline the inner planes,} IP$_1$ (F\MakeLowercase{ig.} S8)}

\begin{figure*}[h]
\includegraphics[width=6.5in]{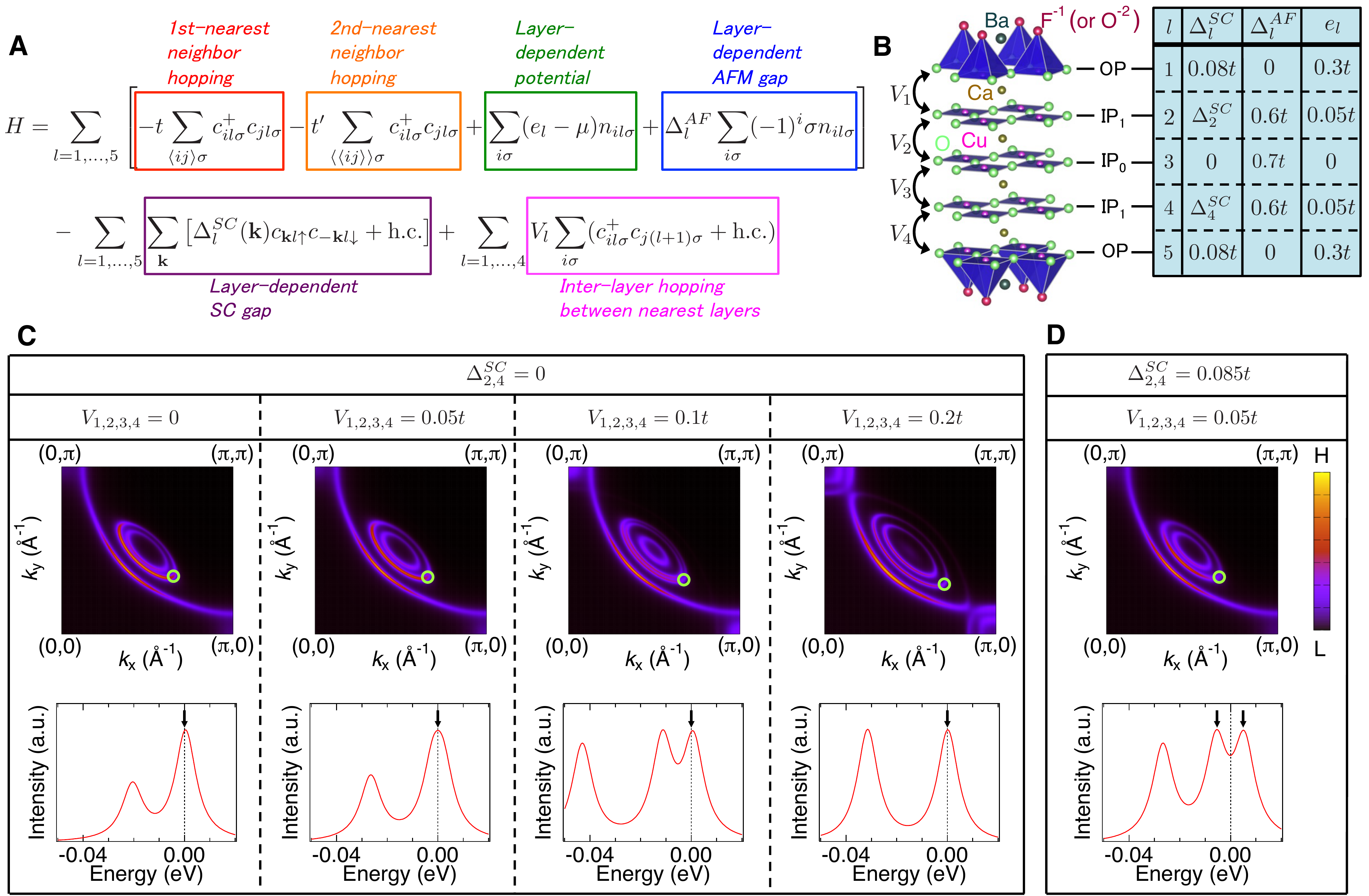}
\caption{{\bf Model calculations demonstrating the superconductivity intrinsic to the inner planes, IP$_1$.}
({\bf A}) Model Hamiltonian expressing the band structure of a system with five CuO$_2$ layers per unit cell; $t=0.14$ eV and $t'=-0.45t$ are found to reproduce well the overall spectral structure of the ARPES results.
({\bf B}) The five CuO$_2$ layers indicated by $l=1-5$: the outer planes (OP), the second inner planes (IP$_1$), and the innermost plane (IP$_0$) are indicated by ($l=1$ and 5),  ($l=2$ and 4), and ($l=$3), respectively. Interlayer hopping between the nearest layers $V_l$ ($l=1-4$) is denoted by arrows.
({\bf C and D}) Fermi surface map obtained by the model Hamiltonian in (A).  
The layer-dependent values of  potential $e _l$, superconducting gap  $\Delta _l^{SC}$, and  antiferromagnetic gap $\Delta _l^{AF}$ are set as listed in the table of (B). From left to right in (C), the interlayer hopping $V_l$ is increased from zero up to 0.2$t$, while  $\Delta _{2,4}^{SC}$ for IP$_1$ is kept to be zero. In (D), $\Delta _{2,4}^{SC}$ and $V_{1,2,3,4}$ are set to be 0.085$t$ and  0.05$t$, respectively. 
The bottom of each Fermi surface map plots the spectrum at $k_F$ of the Fermi pocket for the inner planes IP$_1$ (denoted by light-green circles in the top panels). The arrows point to the peaks of spectrum for IP$_1$; an energy gap is opened in the spectrum of (D), but not in any of the spectra in (C). }
\label{calc}
\end{figure*}

Our ARPES data for inner planes IP$_1$ indicate the coexistence of the superconducting state and  the AF order in the same CuO$_2$ sheet. This is further supported by the model calculation with Hamiltonian shown in Fig. \ref{calc}A, expressing the band structure for a system with five CuO$_2$ layers per unit cell indicated by $l=1-5$; physical meaning for each term is described beside the equation. 
We demonstrate below, in particular, that the superconductivity achieved in IP$_1$ (corresponding to $l=$ 2 and 4 in Fig. \ref{calc}B) originates from itself rather than the superconducting proximity effect from the outer sheet (OP).

The overall spectral structure of the ARPES results is well reproduced by 
$t=0.14$ eV and $t'=-0.45t$. To examine the superconducting proximity effect on the inner layers from the outer layers, we set the layer-dependent values of potential $e _l$, superconducting gap  $\Delta _l^{SC}$, and  antiferromagnetic gap $\Delta _l^{AF}$ as listed in Fig. \ref{calc}B; $\Delta _{2,4}^{SC}$ for  IP$_1$  is not fixed and changed following the two different situations represented in Figs. \ref{calc}C and \ref{calc}D.
 In Fig. \ref{calc}C, $\Delta _{2,4}^{SC}$ for  IP$_1$ is set to be zero, and the interlayer hopping $V_l$ is increased from zero in the left panel up to 0.2$t$ in the right panel. In the bottom of each image, we extract the energy distribution curve (EDC) at $k_F$ of the Fermi pocket for IP$_1$ (denoted by light-green circles in top panels). Importantly, the EDC does not show a superconducting gap (which should be electron-hole symmetric) as indicated by an arrow on the spectral peak even at $V_l=0.2t$, which is so large that FS shows an additional multilayer splitting, unlike the experimental spectra. These results demonstrate that the superconducting proximity effect from OP with $\Delta _{1,5}^{SC}$=0.08$t$ set to be compatible with experiment (see Fig. \ref{calc}B) is not the origin of the observed superconducting gap in IP$_1$. A gap consistent with ARPES results is obtained only when we introduce a finite $\Delta_{2,4}^{SC}$ of 0.085$t$ for IP$_1$, as demonstrated in Fig. \ref{calc}D. Therefore, the spectral gap experimentally observed is attributed to the superconductivity intrinsic to the inner layers IP$_1$, which thus coexists with the AF order in the same CuO$_2$ sheet. 
\\

\section {C\MakeLowercase{omparison of superconducting gaps among  three} F\MakeLowercase{ermi surface sheets \newline for} OP, IP$_1$, \MakeLowercase{and} IP$_0$ (F\MakeLowercase{ig.} S9)}

In the main paper, we present the superconducting gap only for the second inner planes (IP$_1$).
To fully understand the superconducting properties in the five-layered cuprates, we examine here the results also for the outer and innermost planes (OP and IP$_0$, respectively), and compare them with that for IP$_1$. 

\begin{figure*}[h]
\includegraphics[width=6.5in]{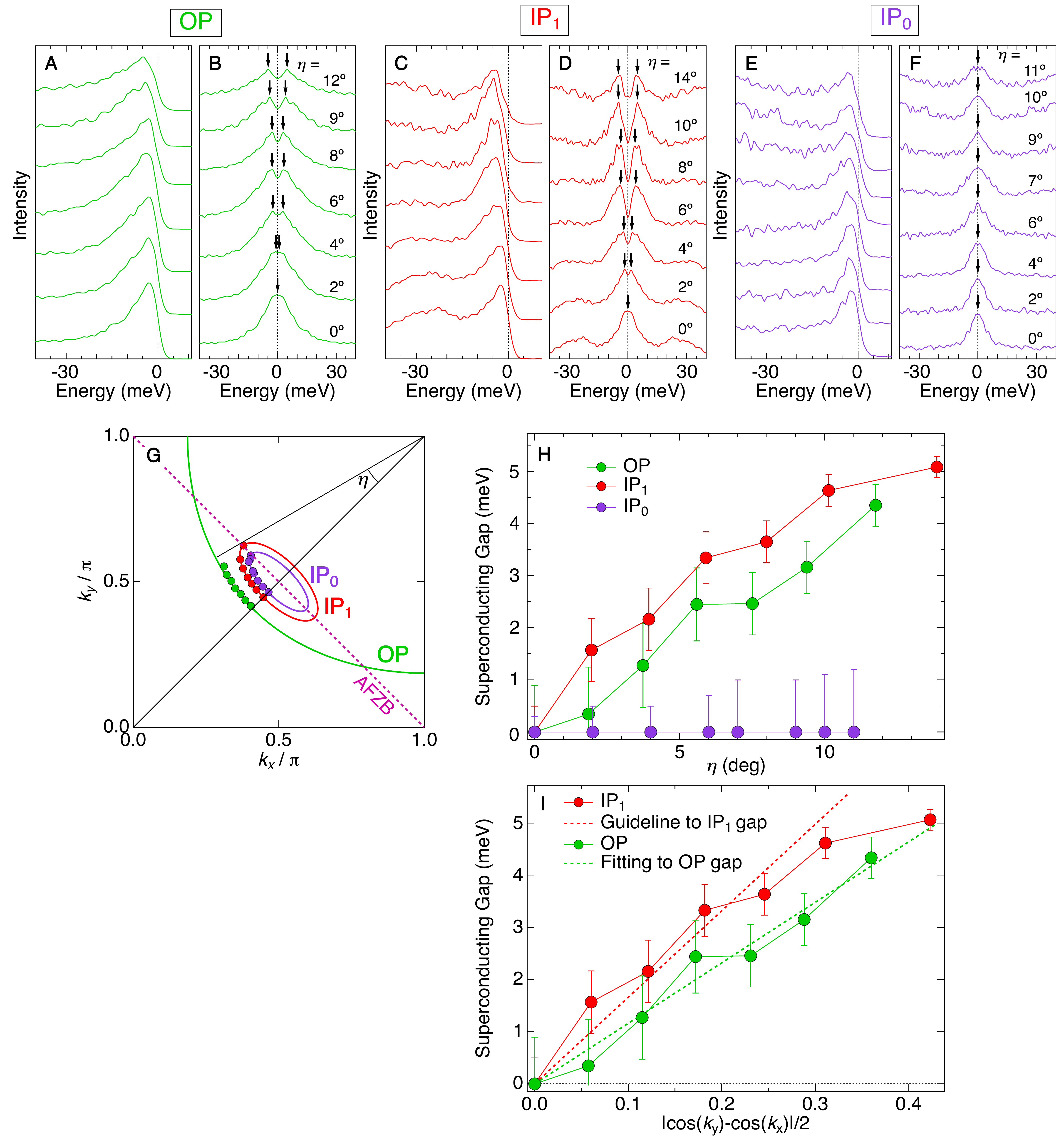}
\caption{{\bf Comparison of superconducting gaps among three Fermi surface sheets for OP, IP$_1$, and IP$_0$.}
({\bf A and B}) Energy distribution curves (EDCs) and those symmetrized about $E_F$, respectively, for the outer plane OP (Fermi arc), measured at $k_F$'s around the gap node (green circles in (G)). The measured angle $\eta$,  defined in (G), is described for each spectrum in the panel. The arrows in (B) point to the peak positions of spectra.  ({\bf C and D}) The same data as in (A and B), but for the second inner plane IP$_1$ (larger Fermi pocket); the measured $k_F$'s are marked by red circles in (G). 
({\bf E and F}) The same data as in (A and B), but for the innermost plane IP$_0$ (smaller Fermi pocket); the measured $k_F$'s are marked by purple circles in (G).
({\bf G}) Fermi surfaces determined by tight-binding fit to our ARPES data for OP, IP$_1$, and IP$_0$. The $k_F$ points measured for the superconducting gap estimation are marked with colored circles.
({\bf H}) The angle  $\eta$ dependence of superconducting gaps estimated from the energies of spectral peaks in (B), (D), and (F). 
({\bf I}) The same data as in (H) for OP and IP$_1$, but plotted as a function of  $\left| {\cos ({k_y})-\cos ({k_x})} \right|/2$.  The data of OP fit well to a straight line (green dashed line), whereas those of IP$_1$ deviate from it (red dashed line).}
\label{gap}
\end{figure*}

Figure \ref{gap}A, \ref{gap}C, and \ref{gap}E plot energy distribution curves (EDCs) measured at $k_F$'s along the Fermi surface for OP (Fermi arc), IP$_1$ (larger Fermi pocket), and  IP$_0$ (smaller Fermi pocket) around the gap node (green, red, and purple circles in Fig. \ref{gap}G, respectively). In Figs. \ref{gap}B, \ref{gap}D, and \ref{gap}F, these spectra are symmetrized about the Fermi level to visualize a gap opening (or gap closing) by eliminating the effect of the Fermi cut-off. In OP (Fig. \ref{gap}B) and IP$_1$ (Fig. \ref{gap}D), the $d$-wave-like gap opens as traced by arrows pointing to the peak positions of spectra: the spectral gap is zero at $\eta$=0$^\circ$ ($\eta$ is defined in Fig. \ref{gap}G), and it opens off the gap node, increasing toward larger $\eta$s. In contrast, we observe no gap all around the Fermi surface (or Fermi pocket) for IP$_0$ (see Fig. \ref{gap}F) within the experimental energy resolution. 

The superconducting gaps estimated are summarized in Fig. \ref{gap}H. In Fig. S9I, we also plot the same results for OP and IP$_1$ as a function of  $\left| {\cos ({k_y})-\cos ({k_x})} \right|/2$. While the gaps for OP with the Fermi arc centered at ($\pi$,$\pi$) is well fitted by a straight line indicating a pure $d$-wave symmetry, those along the Fermi pocket of IP$_1$ strongly deviate from it. The mechanism of the unusual $d$-wave gap symmetry for IP$_1$ is currently unknown and requires further researches to be unveiled. Nonetheless, we found mainly two notable features in these data: (1) the superconducting gap in the Fermi pocket for IP$_1$ is larger than that in the Fermi arc for OP, and  (2) no superconducting gap is opened in the Fermi pocket for IP$_0$. Finding (1) implies that the electron pairing gets more stabilized in the Fermi pocket despite less doping than that in the Fermi arc, perhaps because the pocket can avoid competition with other ordered states, which could develop around the zone edge ($\pi$,0). Finding (2) has two implications as follows. First, the two pockets are almost separately contributed by IP$_0$ and IP$_1$ since, otherwise, the mixing of layers would produce superconducting gaps of similar magnitudes for both the pockets. Secondly, the electronic state of IP$_0$ with less doping than IP$_1$ should be situated outside of the $T_c$ dome in the phase diagram.
\\

\section {T\MakeLowercase{ight-binding fitting to the} ARPES \MakeLowercase{data} (T\MakeLowercase{able} S1)}
In Fig. 3E of the main paper, we exhibit the band dispersion with Fermi pocket determined by tight-binding fitting to our ARPES data in Ba$_2$Ca$_4$Cu$_5$O$_{10}$(F,O)$_2$. For the fitting, six tight-binding functions were used, as listed in Table S1; note that these functions are described for the unit cell enlarged by $\sqrt 2  \times \sqrt 2 $ in the antiferromagnetic state, which reduces the size of the Brillouin zone. The fitting coefficients obtained for each function are also summarized in the same table.

\begin{table}[h]
 \caption{Functions used in the tight-binding fitting to our ARPES data, and obtained coefficients.}
\vspace {-0.3cm} 
\begin{center}
  \begin{tabular}{ccc}
    \hline
    \hline    
      \ \ \ \ \ \ \ \ $i$\ \ \ \ \ \ \ \ 
      &$n_i$&\ \ \ \ \ \ \ \ $t_i$(eV)\ \ \ \ \ \ \ \  \\
    \hline
	0&$1$ &-0.725 \\	
	1&$\frac {1}{2}[\cos (k_x - k_y)+\cos (k_x+k_y)]$ &-0.1275 \\
	2&$\cos (k_x - k_y)\cos (k_x+k_y)$ &-0.81\\
	3&$\frac {1}{2}\{\cos [2(k_x - k_y)]+\cos [2(k_x+k_y)]\}$&-0.0025 \\
	4&$\frac {1}{2}\{\cos [2(k_x - k_y)]\cos (k_x+k_y)+\cos (k_x - k_y)\cos [2(k_x+k_y)]\}$&0.195 \\
	5&$\cos [2 (k_x - k_y)]\cos [2(k_x+k_y)]$ &-0.0075 \\
	6&$\frac {1}{2}\{\cos [3(k_x - k_y)]+\cos [3(k_x+k_y)]\}$ &-0.0675 \\
   \hline
   \hline
 \label {function}        
\end{tabular}
\end{center}
\end{table}

\clearpage

\clearpage

\clearpage

\clearpage

\clearpage

\clearpage

\clearpage

\clearpage


\clearpage

\clearpage




